\pdfoutput=1
\PassOptionsToPackage{dvipsnames}{xcolor}

\documentclass[11pt]{article}

\usepackage[final]{acl}

\usepackage{times}
\usepackage{latexsym}

\usepackage[T1]{fontenc}

\usepackage{microtype}

\usepackage{inconsolata}

\usepackage[utf8]{inputenc}
\usepackage{booktabs,multirow}
\usepackage{enumitem}
\usepackage{kotex} 
\usepackage{color}
\usepackage{rotating}
\usepackage{amsmath}
\usepackage{amsfonts}
\usepackage{pifont}
\usepackage{array}
\usepackage{amssymb}
\usepackage{algpseudocode}
\usepackage{algorithm}
\usepackage{float}
\usepackage{tcolorbox}
\usepackage{multicol}
\usepackage{lipsum}
\usepackage[normalem]{ulem}
\usepackage{xcolor}
\usepackage{soul}
\usepackage{listings}
\usepackage{amsmath} 
\usepackage{adjustbox}
\usepackage{graphicx} 
\usepackage{colortbl} 

\definecolor{customhighlight}{HTML}{DDB6E4}
\sethlcolor{customhighlight}

\colorlet{instructionbg}{gray!15}
\colorlet{questionbg}{gray!25}

\algrenewcommand\algorithmicrequire{\textbf{Input:}}
\algrenewcommand\algorithmicensure{\textbf{Output:}}

\newcommand{\cmark}{\ding{51}}
\newcommand{\xmark}{\ding{55}}

\newcommand{\ie}{{\it i.e.}}
\newcommand{\eg}{{\it e.g.}}

\newcommand{\argmax}{\operatornamewithlimits{argmax}}

\newcommand{\ours}{G{\small RAM}}

\title{GRAM: Generative Recommendation via Semantic-aware\\Multi-granular Late Fusion}

\author{
    Sunkyung Lee$^1$, 
    Minjin Choi$^2$, 
    Eunseong Choi$^1$, 
    Hye-young Kim$^1$, 
    Jongwuk Lee$^1$\thanks{\ \ Corresponding author} \\
    $^1$Sungkyunkwan University, Republic of Korea, 
    $^2$Samsung Research, Republic of Korea \\  
    $^1$\texttt{\{sk1027, eunseong, khyaa3966, jongwuklee\}@skku.edu}, $^2$\texttt{min\_jin.choi@samsung.com}
}

\begin{document}
\maketitle
\begin{abstract}

Generative recommendation is an emerging paradigm that leverages the extensive knowledge of large language models by formulating recommendations into a text-to-text generation task. 
However, existing studies face two key limitations in (i) incorporating implicit item relationships and (ii) utilizing rich yet lengthy item information.
To address these challenges, we propose a \emph{\textbf{G}enerative \textbf{R}ecommender via semantic-\textbf{A}ware \textbf{M}ulti-granular late fusion (\textbf{\ours})}, introducing two synergistic innovations. First, we design \textit{semantic-to-lexical translation} to encode implicit hierarchical and collaborative item relationships into the vocabulary space of LLMs. Second, we present \textit{multi-granular late fusion} to integrate rich semantics efficiently with minimal information loss. It employs separate encoders for multi-granular prompts, delaying the fusion until the decoding stage. 
Experiments on four benchmark datasets show that \ours~outperforms eight state-of-the-art generative recommendation models, achieving significant improvements of 11.5--16.0\% in Recall@5 and 5.3--13.6\% in NDCG@5. The source code is available at \url{https://github.com/skleee/GRAM}.

\end{abstract}

 \section{Introduction}\label{sec:introduction}

Generative recommendation has marked a pivotal shift in recommendation systems, driven by recent advances in large language models (LLMs)~\cite{DevlinCLT19BERT, nips/BrownMRSKDNSSAA20/GPT-3}. While the traditional recommendation approach focuses on matching user and item embeddings within a ranking framework~\cite{KangM18SASRec, ZhouWZZWZWW20S3Rec}, generative recommendation formulates it as a text-to-text generation task~\cite{RajputMSKVHHT0S23TIGER, Geng0FGZ22P5}. It aims to directly generate an \emph{item identifier (ID)} based on the user's historical item sequence.

\begin{figure}
\centering
\includegraphics[width=1.0\linewidth]{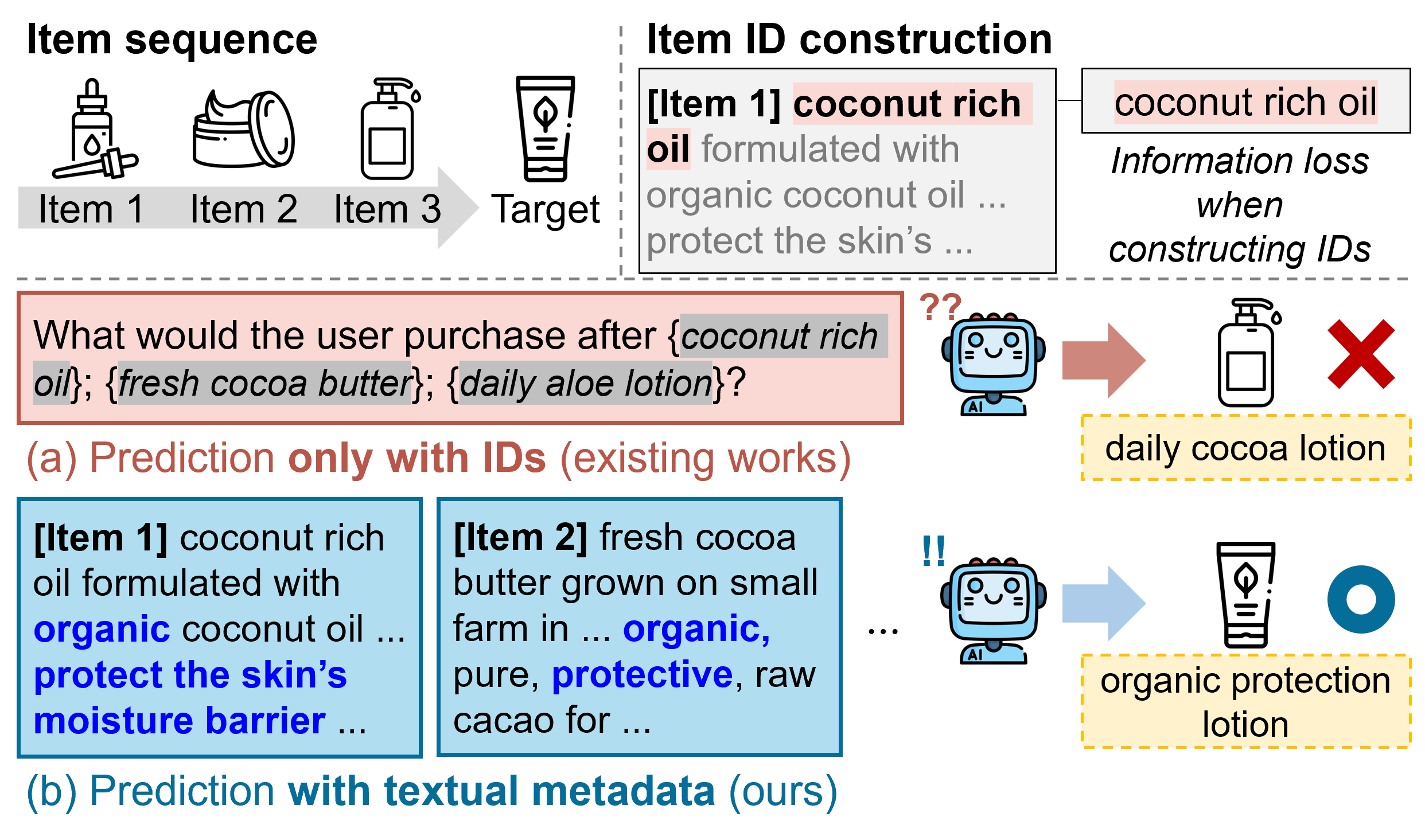}
\caption{Illustration of our motivation. While (a) existing works rely solely on item IDs for prediction, (b) \ours~directly leverages rich textual metadata during prediction, enabling more accurate recommendations.}\label{fig:motivation}
\vspace{-4.5mm}
\end{figure}

A key factor in generative recommendation lies in how well LLMs understand and effectively utilize each item. Existing works~\cite{Geng0FGZ22P5, Tan24IDGenRec, ZhengHLCZCW24LCRec} primarily use rich item metadata to construct abbreviated item IDs, leading to a potential loss of valuable details (Figure~\ref{fig:motivation}a). This limitation motivates us to incorporate item information throughout the entire recommendation process (Figure \ref{fig:motivation}b).

However, two significant challenges hinder LLMs from effectively understanding and utilizing item information.

(i) \emph{How do we enable LLMs to capture implicit item relationships?} While LLMs excel at understanding general language semantics, they struggle with recommendation-specific semantics, \eg, implicit relationships between items~\cite{ZhengHLCZCW24LCRec}. This challenge manifests in two key aspects. 

\begin{figure}
\centering
\includegraphics[width=1.0\linewidth]{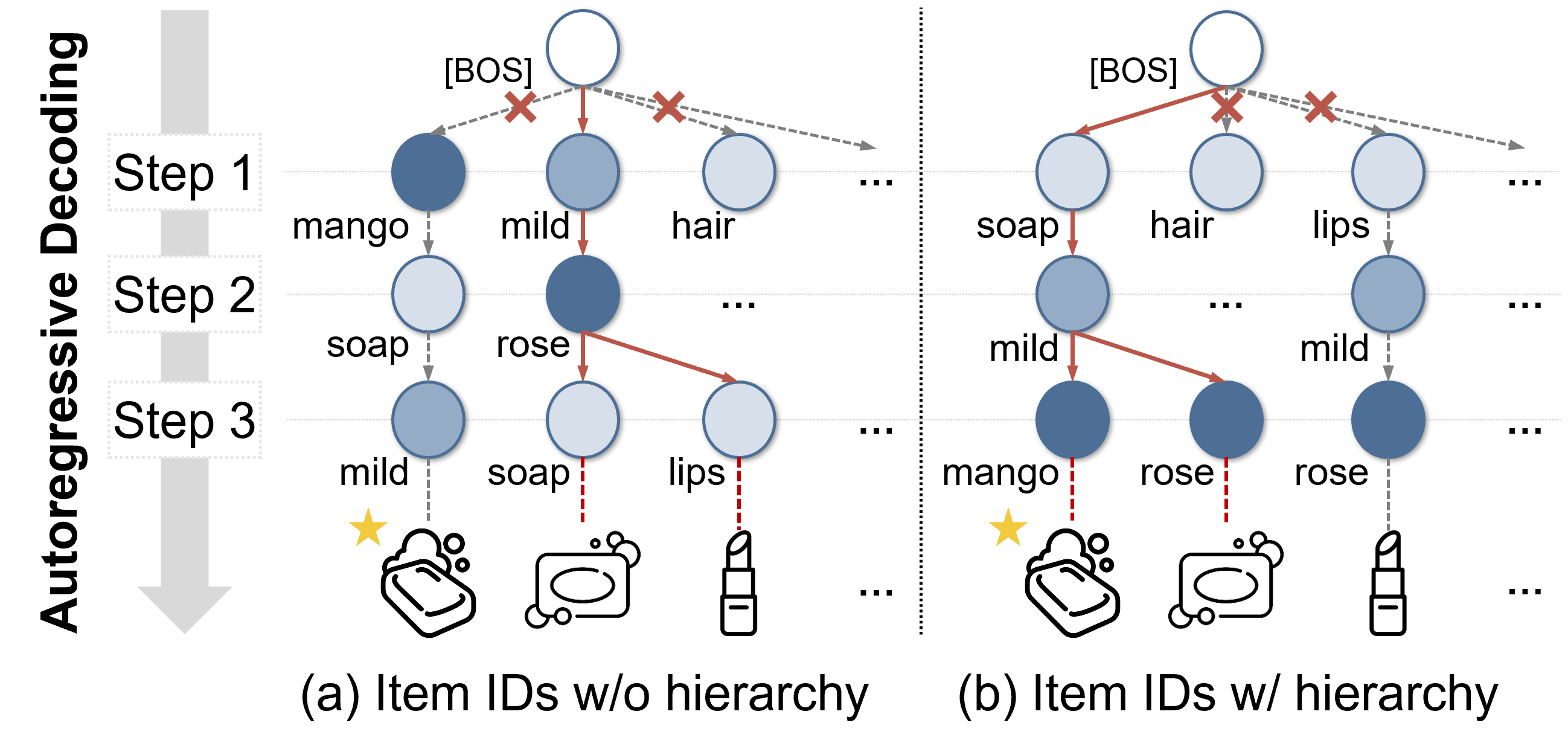}
\vspace{-3mm}
\caption{Illustration of the hierarchy when autoregressively decoding IDs. Darker shades represent more fine-grained information, and targets are marked with stars.}\label{fig:hierarchy}
\vspace{-4.5mm}
\end{figure}

\noindent
\textit{Hierarchical semantics}: Representing the conceptual hierarchy among items (\eg, product taxonomy) is important for accurate and consistent recommendations. Without explicitly encoding the hierarchy in item IDs (Figure~\ref{fig:hierarchy}a), autoregressive generation can lead to semantically inconsistent recommendations, associating unrelated items (`\textit{lipstick}' with `\textit{soap}') based on superficial token-level similarities. In contrast, hierarchical item IDs (Figure~\ref{fig:hierarchy}b) enable semantically coherent recommendations by leveraging shared concepts or attributes. Existing methods using predefined categories~\cite{HuaXGZ23P5Howtoindex} or quantization~\cite{RajputMSKVHHT0S23TIGER} for hierarchical IDs often fail to distinguish similar items or introduce out-of-vocabulary tokens, limiting the LLM's use of pre-trained knowledge.

\noindent
\textit{Collaborative semantics}: Capturing complex user-item interaction patterns is also critical~\cite{ZhengHLCZCW24LCRec}. These collaborative filtering patterns cannot be inferred from a single user sequence alone. While recent work~\cite{ZhengHLCZCW24LCRec} attempts to address this through additional training tasks, it requires extensive fine-tuning to align newly defined IDs with language semantics.

(ii) \emph{How do we handle lengthy item information?} Since items contain rich yet lengthy textual information (\eg, product titles, categories, and descriptions), representing a user history as a sequence of detailed item information leads to excessively long sequences~\cite{SurveyGenSearchRecSys}.\footnote{For the Amazon Beauty dataset~\cite{sigir/McAuleyTSH15Amazonreview}, user sequences consist of 1,440 tokens on average when represented as a simple concatenation of item texts using the T5 tokenizer. Only 4.9\% of sequences have less than 512 tokens.} This poses significant computational challenges due to the quadratic complexity inherent in Transformer-based models. To avoid this, existing studies use partial attributes~\cite{kdd24TransRec} or extract keywords~\cite{Tan24IDGenRec}, inevitably losing information.

To address these challenges, we propose a \emph{\textbf{G}enerative \textbf{R}ecommender via semantic-\textbf{A}ware \textbf{M}ulti-granular late fusion (\textbf{\ours})}, which unlocks the capabilities of LLMs with two key components designed to work synergistically.

\begin{figure}
\centering
\includegraphics[width=1.0\linewidth]{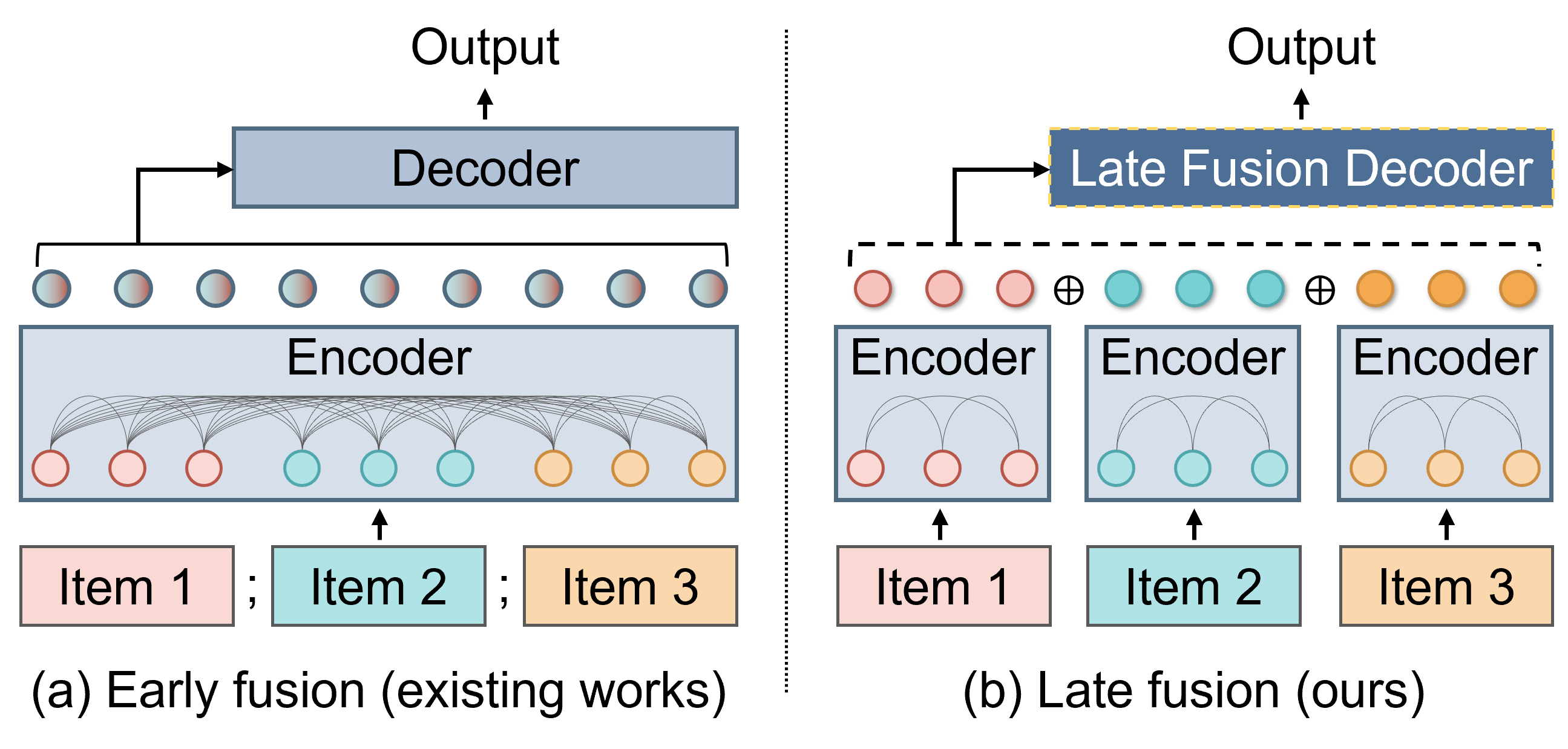}
\vspace{-3mm}
\caption{Schematic diagrams of fusion types. `Item' indicates the textual information. $;$ and $\oplus$ denote the concatenation of text and hidden representations.}\label{fig:fusion}
\vspace{-4.5mm}
\end{figure}

\noindent
(i) \textbf{Semantic-to-lexical translation}. To enable LLM to capture implicit item relationships, we encode item relationships into textual representations prior to training. First, \textit{hierarchical semantics indexing} iteratively clusters the item embeddings based on semantic similarity and maps them into LLM's vocabulary space to generate hierarchical textual IDs. Next, \textit{collaborative semantics verbalization} incorporates collaborative semantics by leveraging a collaborative filtering model. For each item, we identify top-$k$ similar items and express them in a textual format using the item IDs.

\noindent
(ii) \textbf{Multi-granular late fusion}. To effectively handle rich yet lengthy item metadata, we process user history as multiple prompts with different granularities: coarse-grained user prompts for holistic user preferences and fine-grained item prompts for detailed item attributes. Subsequently, multi-granular prompts are efficiently integrated via late fusion. Unlike early fusion (Figure~\ref{fig:fusion}a), which suffers from quadratic complexity due to concatenated texts at the input level, late fusion (Figure~\ref{fig:fusion}b) delays integration until the decoding stage. It aligns with successful techniques in other domains~\cite{eacl/IzacardG21/FiD, YeBPRH23FiD-ICL}. Most importantly, late fusion maximizes the effectiveness of the semantic-to-lexical translation by preserving rich semantics and enabling the processing of significantly longer inputs with minimal information loss.

Our contributions are summarized as follows:

\noindent
\textbf{(i) Item relationship modeling}: We design \textit{semantic-to-lexical translation} to represent item relationships in the vocabulary space of LLMs.

\noindent
\textbf{(ii) Model architecture}: Our \emph{multi-granular late fusion} effectively leverages rich textual item information without expensive computational overhead.

\noindent
\textbf{(iii) Extensive experiments}: \ours~achieves up to 16.0\% and 13.6\% gains in Recall@5 and NDCG@5 over state-of-the-art generative recommenders across four benchmark datasets.

\begin{figure*}
\centering
\includegraphics[width=1.0\linewidth]{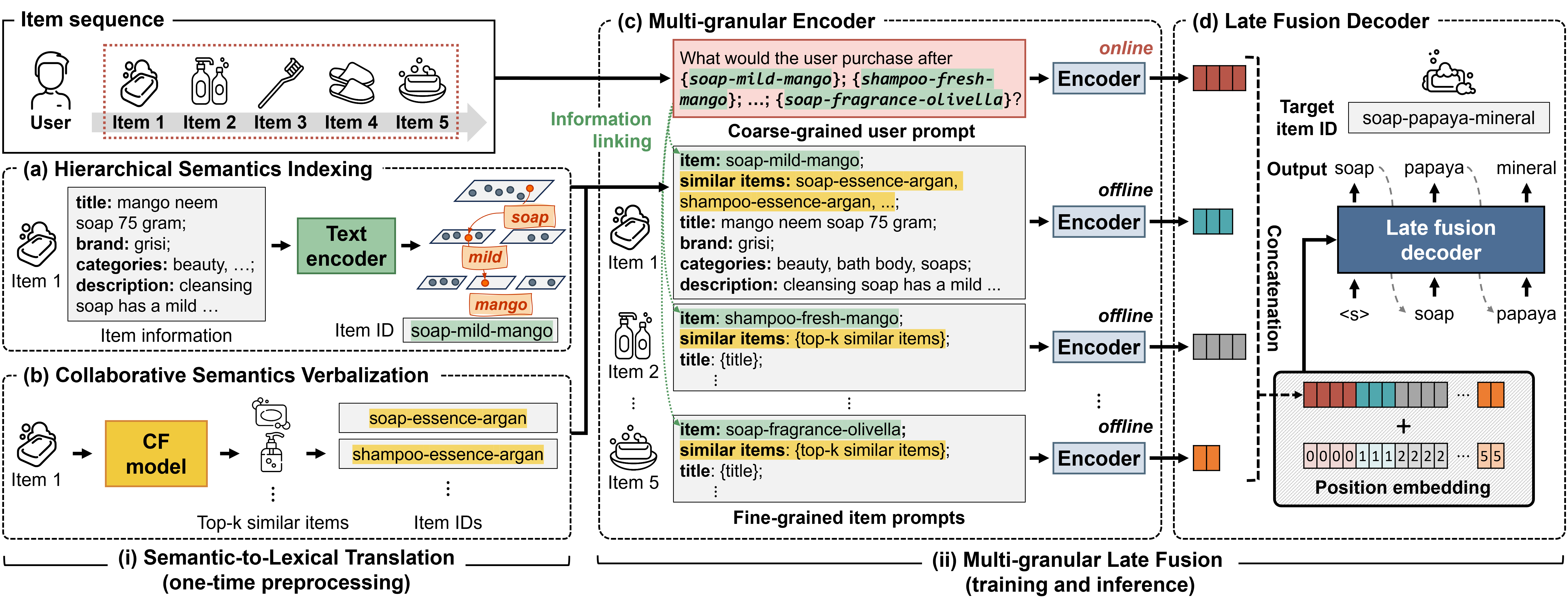}
\vspace{-5.5mm}
\caption{Overall architecture of \ours. (i) We represent hierarchical and collaborative semantics in textual forms via \textit{semantic-to-lexical translation} before training. (ii) The user/item prompts are constructed and encoded separately. The target item ID is inferred via \textit{multi-granular late fusion}, directly leveraging rich textual information.}\label{fig:model}
\vspace{-2.5mm}
\end{figure*}

\section{Related Work}

\subsection{Sequential Recommendation}
It aims to predict the next item based on the user's historical item interactions~\cite{WangHWCSO19SeqSurvey}. To capture user preferences,
item sequences have been modeled with various neural encoders, \eg, RNNs~\cite{HidasiKBT15GRU4Rec}, GNNs~\cite{WuT0WXT19SRGNN}, and Transformers~\cite{KangM18SASRec, SunLWPLOJ19BERT4Rec,kdd/MaKL19HGN}. Recently, several studies~\cite{ZhangZLSXWLZ19FDSA, ZhouWZZWZWW20S3Rec} have used metadata to model item dependency and user-item interactions. However, they express attributes as discrete IDs, neglecting to exploit \textit{textual metadata} with LLMs.

\subsection{Generative Recommendation}

Given an item sequence, it generates a \textit{target item identifier}.\footnote{For reranking~\cite{corr/abs-2303-14524/ChatREC} or discriminative methods~\cite{HouMZLDW22UniSRec}, refer to Appendix~\ref{sec:app_relatedwork}.} Existing studies can be categorized depending on the type of item identifiers.

\noindent
\textbf{Numeric IDs}\footnote{While \citet{cikm24LETTER, SurveyGenSearchRecSys} distinguish codebook and numeric IDs, we group them as numeric IDs.}. 
P5~\cite{Geng0FGZ22P5, HuaXGZ23P5Howtoindex} adopts numeric IDs and multi-task learning. TIGER~\cite{RajputMSKVHHT0S23TIGER} and LC-Rec~\cite{ZhengHLCZCW24LCRec} construct codebooks with vector quantization, \ie, RQ-VAE~\cite{ZeghidourLOST22RQVAE}. LC-Rec further adopted alignment tasks for language and collaborative semantics. LETTER~\cite{cikm24LETTER} improved vector quantization by integrating collaborative signals. Recently, ELMRec~\cite{WangCFS24ELMRec} incorporates high-order relationships from the graph. However, since numeric IDs are separated from the vocabulary of LLMs, they suffer from a semantic gap that hinders the full potential of LLMs for recommendations.

\begin{table}[]\footnotesize
\centering
\setlength{\tabcolsep}{2.2pt}
\renewcommand{\arraystretch}{0.9}
\begin{tabular}{>{\centering\arraybackslash}p{1.4cm}|>{\centering\arraybackslash}p{2cm} >{\centering\arraybackslash}p{1.3cm}|>{\centering\arraybackslash}p{1cm}>{\centering\arraybackslash}p{1.2cm}}
\toprule
\multirow{2}{*}{Model} & \multicolumn{2}{c|}{Textual metadata usage} & \multicolumn{2}{c}{Item ID} \\
 & ID construction & Prediction & Text & Hierarchy \\ \midrule
P5-SemID & \textcolor{Green}{\cmark} & \textcolor{Red}{\xmark} & \textcolor{Red}{\xmark} & \textcolor{Green}{\cmark} \\
TIGER & \textcolor{Green}{\cmark} & \textcolor{Red}{\xmark} & \textcolor{Red}{\xmark} & \textcolor{Green}{\cmark} \\ 
IDGenRec & \textcolor{Green}{\cmark} & \textcolor{Red}{\xmark} & \textcolor{Green}{\cmark} & \textcolor{Red}{\xmark} \\
LETTER & \textcolor{Green}{\cmark} & \textcolor{Red}{\xmark} & \textcolor{Red}{\xmark} & \textcolor{Green}{\cmark} \\
ELMRec & \textcolor{Red}{\xmark} & \textcolor{Red}{\xmark} & \textcolor{Red}{\xmark} & \textcolor{Red}{\xmark} \\
LC-Rec & \textcolor{Green}{\cmark} & \textcolor{Red}{\xmark} & \textcolor{Red}{\xmark} & \textcolor{Green}{\cmark} \\ \midrule
\textbf{G{\scriptsize RAM}} & \textbf{\textcolor{Green}{\cmark}} & \textbf{\textcolor{Green}{\cmark}} & \textcolor{Green}{\cmark} & \textbf{\textcolor{Green}{\cmark}} \\ 
\bottomrule
\end{tabular}
\caption{Comparison of different generative recommendation models on (i) how textual metadata is utilized and (ii) how item IDs are constructed.}\label{tab:app_comparison}
\vspace{-4.5mm}
\end{table}

\noindent
\textbf{Textual IDs}. 
Another line of research has explored textual IDs as a meaningful alternative to numeric IDs. IDGenRec~\cite{Tan24IDGenRec} generates semantic item IDs from text metadata using an ID generator. TransRec~\cite{kdd24TransRec} combines both numeric and textual IDs in a hybrid approach. However, none of them explore how to (i) directly utilize textual metadata during prediction and (ii) incorporate the item hierarchy into the textual identifiers. Further comparison is presented in Table~\ref{tab:app_comparison}.

\section{Proposed Method} \label{sec:method}

We present a \emph{\textbf{G}enerative \textbf{R}ecommender via semantic-\textbf{A}ware \textbf{M}ulti-granular late fusion (\textbf{\ours})}, as depicted in Figure~\ref{fig:model}. \ours~seamlessly incorporates item relationships (Section~\ref{sec:sem_to_lex}) and fully utilizes rich item information (Section~\ref{sec:fusion}). We lastly explain the training and inference processes of \ours~(Section~\ref{sec:train_inference}).

\subsection{Task Formulation}\label{sec:model_problem}

Let $\mathcal{U}$ and $\mathcal{I}$ denote a set of users and items. For each user $u \in \mathcal{U}$, we represent the interaction history as a chronological item sequence $s_u=(i_1, \dots, i_{|s|})$, where $i_t$ is the item at the $t$-th position, and $|s|$ indicates the number of items in the sequence $s$. The goal is to predict the next item $i_{|s|+1}$ that the user is most likely to interact with based on the user's interaction history.

Each item $i \in \mathcal{I}$ is assigned to the unique ID $\Tilde{i}$. Each user's item sequence $s_u$ is converted to the sequence of item IDs, \ie, $\Tilde{s}_u=(\Tilde{i}_1, \Tilde{i}_2, \dots, \Tilde{i}_{|s|})$. Given the sequence $\Tilde{s}_u$, the generative recommendation is formulated by generating the next item ID $\Tilde{i}_{|s|+1}$, which the user is most likely to prefer.

Each item consists of multiple attributes, denoted by $(a_1, \dots, a_m)$, where $m$ is the number of attributes. Each attribute indexed by $j$ is presented in a key-value format $a_j= (k_j, v_j)$, where $k_j$ is the attribute name $a_j$, \eg, ``\textit{title},'' ``\textit{brand},'' or ``\textit{description},'' and $v_j$ represents the corresponding attribute value. We represent item text by combining the attributes, \eg, ``\textit{title: coastal scents cocoa butter; brand: coastal scents; description: ...}''.

\begin{figure}
\centering
\includegraphics[width=1.0\linewidth]{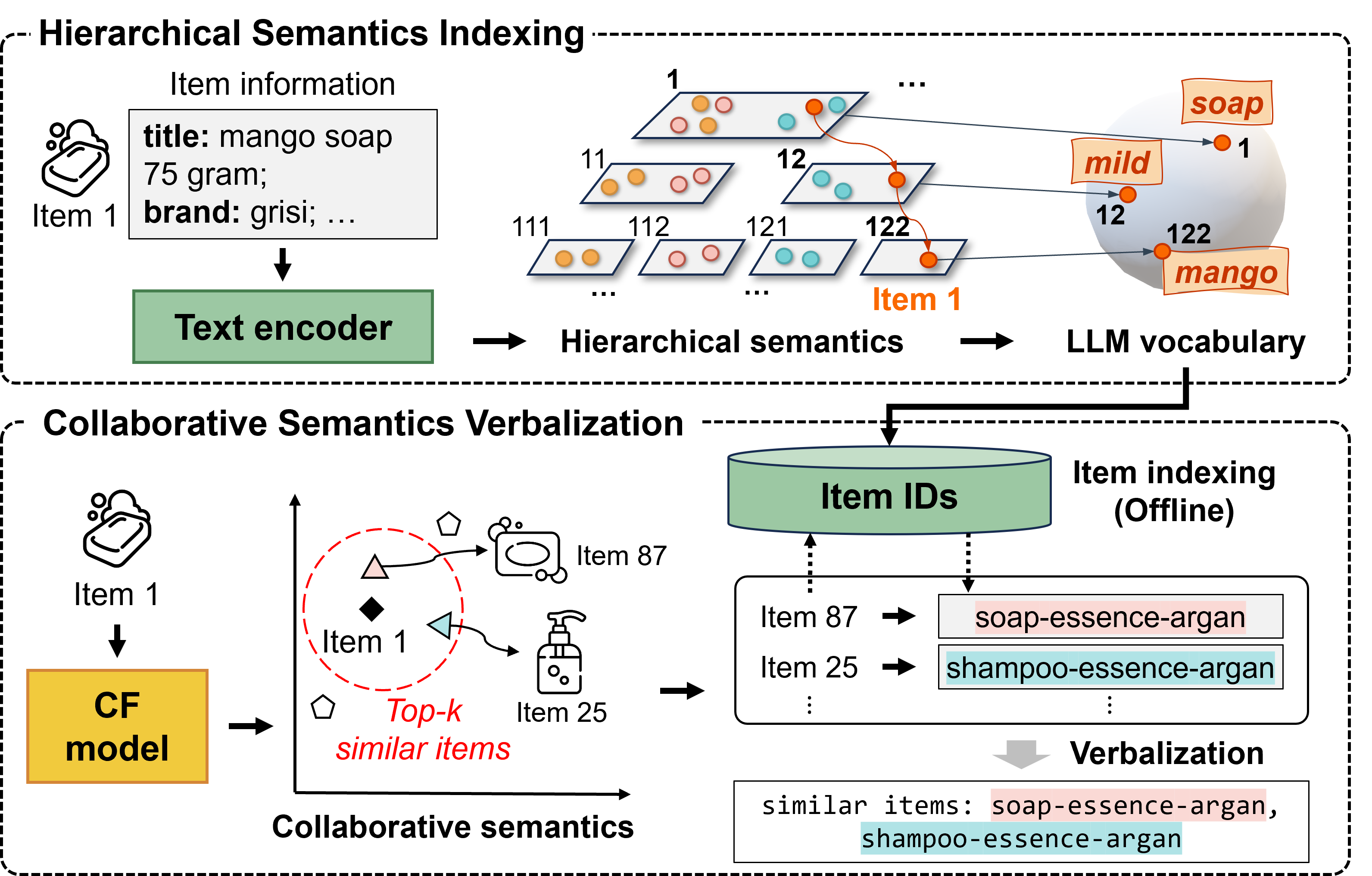}
\caption{Illustration of semantic-to-lexical translation. Each item is assigned textual IDs based on \textit{hierarchical semantics}, and \textit{collaborative semantics} are verbalized based on the hierarchical IDs.}\label{fig:cf_hid}
\vspace{-4.5mm}
\end{figure}

\subsection{Semantic-to-Lexical Translation}\label{sec:sem_to_lex}

We introduce semantic-to-lexical translation, representing implicit item relationships in a textual form. Figure~\ref{fig:cf_hid} depicts \textit{hierarchical semantics indexing} that transforms the item hierarchy into textual IDs, and \textit{collaborative semantics verbalization} that represents collaborative signals as extra attributes using textual IDs. Note that these are preprocessing steps performed only once before training.

\subsubsection{Hierarchical Semantics Indexing}\label{sec:item_id_indexing}

We present a novel method for constructing hierarchical textual IDs, offering three key advantages: (i) utilizing LLMs' knowledge through natural language tokens, (ii) capturing semantic relationships among items through shared identifier prefixes, and (iii) enabling the progressive generation of identifiers from general to specific item characteristics during autoregressive decoding~\cite{nips/Tay/DSI}.

\noindent
\textbf{Hierarchical Semantics Extraction}. 
We employ hierarchical $k$-means clustering over item embeddings to construct hierarchical identifiers where semantically similar items have identical prefixes. Each item embedding $\textbf{z} \in \mathbb{R}^{e}$ is obtained using the text encoder. The clustering process begins by partitioning all items into $k$ clusters. For clusters with more than $c$ items, we recursively apply the $k$-means clustering to further divide it into sub-clusters. The recursive clustering process terminates when the cluster size is smaller than $c$ or the maximum depth $l$ is reached. Finally, each item is assigned a sequence of cluster indices, where the sequence length is bounded by $l$.

\noindent
\textbf{Hierarchical Semantics Translation}. To translate hierarchical semantics into LLM vocabulary, we assign representative tokens to each cluster. \ours~deliberately utilizes existing tokens, unlike previous studies~\cite{RajputMSKVHHT0S23TIGER, ZhengHLCZCW24LCRec} that employ out-of-vocabulary tokens. That is, we preserve the hierarchical structure and minimize potential conflicts with language semantics. We convert each text of item $i$ into a $|V|$-dimensional vocabulary space vector $\mathbf{V}_i$. For simplicity, we use the TF-IDF~\cite{Jones2004/TFIDF} scoring schema. Note that more sophisticated scoring functions~\cite{sigir/FormalPC21/SPLADE} can be used. We then create a cluster-level vocabulary vector by averaging vectors for all items within a cluster. Lastly, we select the most representative token, \ie, the token with the highest score, from the cluster-level vector. For items whose cluster indices length is shorter than $l$, we append additional tokens using the vocabulary vector $\mathbf{V}_i$ to ensure a length of $l$. For duplicate IDs, we append an additional digit for uniqueness. (See the detailed algorithm in Appendix~\ref{sec:app_clustering}.)

\subsubsection{Collaborative Semantics Verbalization}
We integrate collaborative semantics into LLM-based recommenders to complement LLM's capabilities. While LLMs excel at processing textual semantics, they still struggle to incorporate collaborative patterns across items~\cite{ZhengHLCZCW24LCRec, kdd24ALLMRec}. To bridge collaborative semantics to LLMs, we extract collaborative signals and convert them as an additional item attribute.

\noindent
\textbf{Collaborative Semantics Extraction}. We employ the off-the-shelf collaborative filtering (CF) model\footnote{We utilized SASRec~\cite{KangM18SASRec}.} to capture item relationships through learned embeddings. We select the top-$k$ most similar items for each item $i$:
\begin{equation}\label{eq:top_k_fetching}
    i^{\ CF}_1, \dots, i^{\ CF}_k = \texttt{argTop-$k$}_{\substack{j \in \mathcal{I}}} \text{sim}(\mathbf{e}_i, \mathbf{e}_j),
\end{equation}
where $\mathbf{e}_i$ and $\mathbf{e}_j$ represent item embedding vectors obtained from the off-the-shelf CF model. The similarity is computed by a function $\text{sim}(\cdot,\cdot)$, \eg, dot product. The $\texttt{argTop-$k$}(\cdot)$ function returns the indices of $k$ items with the highest similarity scores.

\noindent
\textbf{Collaborative Semantics Translation}. 
We then transform the collaborative knowledge into a text format using hierarchical item IDs:
\begin{equation}\label{eq:cf_verbalization}\small
    a_{CF} = (k_{CF}, v_{CF}), \ \ \text{where} \ \ v_{CF} = [ \Tilde{i}^{\ CF}_1, \dots, \Tilde{i}^{\ CF}_k ].
\end{equation}
Here, $k_{CF}$ represents the key `\textit{similar items}', and $v_{CF}$ is the verbalized similar items, \eg, ``\textit{soap-essence-argan, shampoo-essence-argan, ...}''. It allows us to incorporate collaborative signals when using LLMs. It is also well aligned with existing findings~\cite{corr/abs-2311-10779DOKE, acl24BinLLM}.

\subsection{Multi-granular Late Fusion}\label{sec:fusion}
To effectively leverage rich but lengthy item information, we introduce \textit{multi-granular late fusion}. We process inputs into coarse-grained user prompts and fine-grained item prompts with \textit{multi-granular encoder}. The \textit{late fusion decoder} then integrates them while preserving detailed information during the prediction. By delaying fusion until the decoder, it avoids the quadratic complexity and achieves enhanced efficiency, as evidenced by the theoretical and empirical analysis in Appendix~\ref{sec:app_efficiency}.

Importantly, multi-granular late fusion synergistically leverages the item relationships from semantic-to-lexical translation. Hierarchical relationships in item IDs are incorporated into user prompts, while collaborative relationships in item attributes are integrated into item prompts. This design enables \ours~to fully exploit the benefits of semantic-to-lexical translation.

\subsubsection{Multi-granular Encoder}\label{sec:prompt}
To capture user preference, existing methods~\cite{Geng0FGZ22P5, Tan24IDGenRec} combine item IDs with prompt templates. While the item ID sequence provides a holistic view of user behavior and explicitly captures sequential dependencies, it inevitably loses item details. Conversely, using full text preserves details but leads to lengthy inputs. Thus, we adopt two complementary prompts to exploit the distinctive benefits, as exemplified in Appendix~\ref{sec:app_prompt}.

\noindent
\textbf{Coarse-grained User Prompt}.
We capture the overall user preferences through a concise representation of the user interaction history by concatenating item IDs as follows:
    \begin{equation}\label{eq:sequence_long}
    \Tilde{s}_{seq}=[ \Tilde{i}_{|s|} ; \Tilde{i}_{|s|-1} ; \dots ; \Tilde{i}_1 ],
    \end{equation}
where $;$ is the concatenation of text. The user history is sorted in reverse order to prevent recent items from being truncated~\cite{LiWLFSSM23Recformer}. To transform it into a natural language, we interpolate it into the placeholder of a predefined prompt:
\begin{equation}\nonumber\footnotesize
    T_{u}=\texttt{``What would the user purchase after \{$\Tilde{s}_{seq}$\}?''}
\end{equation}

\noindent
\textbf{Fine-grained Item Prompt}.
We represent detailed item characteristics and relationships by leveraging comprehensive item attributes and collaborative semantics $a_{CF}$, which are obtained from Eq.~\eqref{eq:cf_verbalization}. The item prompt $T_{i}$ is constructed as follows:
\begin{equation}\label{eq:id_append}
    T_i = (a_{ID}, a_{CF}, a_1, \dots, a_m),
\end{equation}
where $a_{ID} = (\text{`item:'}, \Tilde{i} \ )$ represents the item identifier such as ``\textit{item: soap-mild-mango}.'' Notably, we append the IDs to link user prompts with their corresponding item prompts.

\noindent
\textbf{Prompt Encoding}.
The multi-granular user and item prompts are processed independently and represented as a list of prompts:
\begin{equation}\label{eq:fid_input}
     \mathcal{P} = ( T_{u}, T_{i_{|s|}}, \dots, T_{i_1} ).
\end{equation}
With one user prompt and $|s|$ item prompts, we encode each prompt $\mathcal{P}_j$ using the T5 encoder~\cite{JMLR/Raffel2020/T5} to obtain token embeddings $\mathbf{H}_j$:
\begin{equation}\label{eq:fid_encoder}\small
    \mathbf{H}_j=\text{Encoder}\left(\mathcal{P}_j\right) \in \mathbb{R}^{M \times d} \ \text{for} \  j \in \{1, \dots, |s|+1\},
\end{equation}
where $M$ is the maximum text sequence length of the encoder, and $d$ is the hidden dimension size. 

It is non-trivial to connect items in the user prompt with their corresponding item prompts. We resolve it through \textit{information linking}, where each item ID ($a_{ID}$) serves as a bridge between coarse- and fine-grained information. This linking effectively integrates information across the prompts.

\subsubsection{Late Fusion Decoder}
The decoder integrates the representations of user preferences and item information to generate recommendations. To explicitly encode the sequential information of user interactions, we incorporate position embedding $\mathbf{P} \in \mathbb{R}^{(L+1) \times d}$, where $L$ is the maximum number of items in user history:
\begin{equation}\label{eq:fid_position}
    \mathbf{X}_j = \mathbf{H}_j + \mathbf{P}_j \ \ \text{for} \ \ j \in \{1, \dots, |s|+1\}.
\end{equation}
We then combine all position-aware representations $\mathbf{X}_j$ into a unified embedding matrix $\mathbf{X}$:
\begin{equation}\label{eq:fid_encoder_output}
    \mathbf{X} = [ \mathbf{X}_1; \dots; \mathbf{X}_{|s|+1} ] \in \mathbb{R}^{( (|s|+1) \times M) \times d} .
\end{equation}
Finally, the decoder leverages this comprehensive representation $\mathbf{X}$ as the key-value matrix in cross-attention to aggregate the rich textual information. The target item ID is generated autoregressively, considering both coarse-grained user preferences and fine-grained item attributes. The probability of generating a textual item ID $\Tilde{i}$ is defined as:
\begin{equation}\label{eq:generate_target_id}
    P(\Tilde{i} | \mathcal{P}) = \prod_{t=1}^{n} P(\Tilde{i}^t | \mathcal{P}, \Tilde{i}^{<t}).
\end{equation}

\subsection{Training and Inference}\label{sec:train_inference}
For training, the model learns to generate textual IDs by minimizing the sequence-to-sequence cross-entropy loss with teacher-forcing:
\begin{equation}\label{eq:seq2seq_loss}
    \mathcal{L} = - \sum_{t=1}^{n} \text{log} P(\Tilde{i}^t | \mathcal{P}, \Tilde{i}^{<t}), \\
\end{equation}
where $\Tilde{i}^{t}$ is a $t$-th token of the target item ID $\Tilde{i}_{|s|+1}$.

For inference, we adopt a two-stage process. In the offline stage (\ie, pre-processing), we assign IDs to all items, obtain collaborative knowledge from the CF model, and pre-compute the encoder outputs for fine-grained item prompts. 
Then, during the online stage, we only encode the user prompt and generate the recommendations using constrained beam search. To generate valid IDs, we use a prefix tree Trie~\cite{cormen2022intro2algoTrie} following~\citet{ZhengHLCZCW24LCRec, emnlp/L23GLEN}. This two-stage approach significantly reduces computational overhead by processing fine-grained item information offline.

\section{Experimental Setup}\label{sec:setup}

\begin{table*}[]\scriptsize
\centering
\setlength{\tabcolsep}{3.32pt}
\renewcommand{\arraystretch}{1.2}
\begin{tabular}{ccccc|cccc|cccc|cccc}
\toprule
\multicolumn{1}{c|}{\multirow{2}{*}{Model}} & \multicolumn{4}{c|}{Beauty} & \multicolumn{4}{c|}{Toys} & \multicolumn{4}{c|}{Sports} & \multicolumn{4}{c}{Yelp} \\
\multicolumn{1}{c|}{} & R@5 & N@5 & R@10 & N@10 & R@5 & N@5 & R@10 & N@10 & R@5 & N@5 & R@10 & N@10 & R@5 & N@5 & R@10 & N@10 \\ \midrule
& \multicolumn{16}{c}{\textit{Traditional recommendation models}} \\ \midrule
\multicolumn{1}{c|}{GRU4Rec} & 0.0429 & 0.0288 & 0.0643 & 0.0357 & 0.0371 & 0.0254 & 0.0549 & 0.0311 & 0.0237 & 0.0154 & 0.0373 & 0.0197 & 0.0240 & 0.0157 & 0.0398 & 0.0207 \\
\multicolumn{1}{c|}{HGN}  & 0.0350 & 0.0217 & 0.0589 & 0.0294 & 0.0345 & 0.0212 & 0.0553 & 0.0279 & 0.0203 & 0.0127 & 0.0340 & 0.0171 & 0.0366 & 0.0250 & 0.0532 & 0.0304 \\
\multicolumn{1}{c|}{SASRec} & 0.0323 & 0.0200 & 0.0475 & 0.0249 & 0.0339 & 0.0208 & 0.0442 & 0.0241 & 0.0147 & 0.0089 & 0.0220 & 0.0113 & 0.0284 & 0.0214 & 0.0353 & 0.0245 \\ 
\multicolumn{1}{c|}{BERT4Rec} & 0.0267 & 0.0165 & 0.0450 & 0.0224 & 0.0210 & 0.0131 & 0.0355 & 0.0178 & 0.0136 & 0.0085 & 0.0233 & 0.0116 & 0.0244 & 0.0159 & 0.0401 & 0.0210 \\
\multicolumn{1}{c|}{FDSA} & \ul{0.0570} & \ul{0.0412} & \ul{0.0777} & \ul{0.0478} & \ul{0.0619} & \ul{0.0455} & \ul{0.0805} & \ul{0.0514} & 0.0283 & 0.0201 & 0.0399 & 0.0238 & 0.0331 & 0.0218 & \ul{0.0534} & 0.0284 \\
\multicolumn{1}{c|}{S$^3$Rec} & 0.0377 & 0.0235 & 0.0627 & 0.0315 & 0.0365 & 0.0231 & 0.0592 & 0.0304 & 0.0229 & 0.0145 & 0.0370 & 0.0190 & 0.0190 & 0.0117 & 0.0321 & 0.0159 \\ \midrule
& \multicolumn{16}{c}{\textit{Generative recommendation models}} \\ \midrule
\multicolumn{1}{c|}{P5-SID} & 0.0465 & 0.0329 & 0.0638 & 0.0384 & 0.0216 & 0.0151 & 0.0325 & 0.0186 & 0.0295 & 0.0212 & 0.0403 & 0.0247 & 0.0299 & 0.0211 & 0.0432 & 0.0253 \\
\multicolumn{1}{c|}{P5-CID} & 0.0465 & 0.0325 & 0.0668 & 0.0391 & 0.0223 & 0.0143 & 0.0357 & 0.0186 & 0.0295 & 0.0214 & 0.0420 & 0.0254 & 0.0226 & 0.0155 & 0.0363 & 0.0199 \\
\multicolumn{1}{c|}{P5-SemID} & 0.0459 & 0.0327 & 0.0667 & 0.0394 & 0.0264 & 0.0178 & 0.0416 & 0.0227 & \ul{0.0336} & \ul{0.0243} & \ul{0.0481} & \ul{0.0290} & 0.0212 & 0.0143 & 0.0329 & 0.0181 \\
\multicolumn{1}{c|}{TIGER} & 0.0352 & 0.0236 & 0.0533 & 0.0294 & 0.0274 & 0.0174 & 0.0438 & 0.0227 & 0.0176 & 0.0143 & 0.0311 & 0.0146 & 0.0164 & 0.0103 & 0.0262 & 0.0135 \\
\multicolumn{1}{c|}{IDGenRec$^\dagger$} & 0.0463 & 0.0328 & 0.0665 & 0.0393 & 0.0462 & 0.0323 & 0.0651 & 0.0383 & 0.0273 & 0.0186 & 0.0403 & 0.0228 & 0.0310 & 0.0219 & 0.0448 & 0.0263 \\ 
\multicolumn{1}{c|}{LETTER} & 0.0364 & 0.0243 & 0.0560 & 0.0306 & 0.0309 & 0.0296 & 0.0493 & 0.0262 & 0.0209 & 0.0136 & 0.0331 & 0.0176 & 0.0298 & 0.0218 & 0.0403 & 0.0252 \\
\multicolumn{1}{c|}{ELMRec$^\dagger$}  & 0.0372 & 0.0267 & 0.0506 & 0.0310 & 0.0148 & 0.0119 & 0.0193 & 0.0131 & 0.0241 & 0.0181 & 0.0307 & 0.0203 & \ul{0.0424} & \ul{0.0301} & 0.0501 & \ul{0.0324} \\
\multicolumn{1}{c|}{LC-Rec} & 0.0503 & 0.0352 & 0.0715 & 0.0420 & 0.0543 & 0.0385 & 0.0753 & 0.0453 & 0.0259 & 0.0175 & 0.0384 & 0.0216 & 0.0341 & 0.0235 & 0.0501 & 0.0286 \\
\midrule
\rowcolor{gray!20} 
\multicolumn{1}{c|}{\textbf{G{\tiny RAM}}}  & \textbf{0.0641} & \textbf{0.0451} & \textbf{0.0890} & \textbf{0.0531} & \textbf{0.0718} & \textbf{0.0516} & \textbf{0.0987} & \textbf{0.0603} & \textbf{0.0375} & \textbf{0.0256} & \textbf{0.0554} & \textbf{0.0314} & \textbf{0.0476} & \textbf{0.0326} & \textbf{0.0698} & \textbf{0.0397} \\ \midrule
\multicolumn{1}{c|}{Gain (\%)}  & 12.4$^*$ & 9.5$^*$ & 14.5$^*$ & 11.0$^*$ & 16.0$^*$ & 13.6$^*$ & 22.7$^*$ & 17.1$^*$ & 11.5$^*$ & 5.3$^*$ & 15.2$^*$ & 8.3$^*$ & 12.3$^*$ & 8.1$^*$ & 30.7$^*$ & 22.5$^*$ \\
\bottomrule
\end{tabular}
\vspace{-1.5mm}
\caption{Overall performance comparison. The best model is marked in \textbf{bold}, and the second-best model is \ul{underlined}. Gain measures improvement of the proposed method over the best competitive baseline. `$*$' indicates statistical significance $(p < 0.05)$  by a paired $t$-test. `$\dagger$' indicates models where our reproduced results differ from the original papers due to corrected experimental settings. Please refer to Appendix~\ref{sec:app_idgenrec} and~\ref{sec:app_elmrec} for details. Efficiency analysis is in Appendix~\ref{sec:app_efficiency}. Additional results with a cutoff of 20 are in Appendix~\ref{sec:app_cutoff20}.}\label{tab:exp_overall}
\end{table*}

\noindent
\textbf{Datasets}. 
We conduct experiments on four real-world datasets: Amazon review~\cite{sigir/McAuleyTSH15Amazonreview, www/HeM16Amazonreview}\footnote{\url{https://jmcauley.ucsd.edu/data/amazon/}} and Yelp\footnote{\url{https://www.yelp.com/dataset}}. Among the Amazon datasets, we select three subcategories: ``Sports and Outdoors'', ``Beauty'', and ``Toys and Games''. Following \citet{HuaXGZ23P5Howtoindex}, we remove users and items with fewer than five interactions (5-core setting). The statistics are in Appendix~\ref{sec:app_dataset}.

\vspace{1mm}
\noindent
\textbf{Evaluation Protocols and Metrics}. 
We employ the \emph{leave-one-out} strategy to split the train, validation, and test sets following \cite{KangM18SASRec,ZhengHLCZCW24LCRec}. For each user sequence, we use the last item as test data, the second last item as validation data, and the remaining items as training data. We conduct \textit{full-ranking evaluations} on all items rather than on sampled items for an accurate assessment. For metrics, we adopt top-$k$ Recall (R@$k$) and Normalized Discounted Cumulative Gain (N@$k$) with cutoff $k=\{5, 10\}$.

\vspace{1mm}
\noindent
\textbf{Baselines}. 
We adopt six traditional sequential recommenders: \textbf{GRU4Rec}~\cite{HidasiKBT15GRU4Rec}, \textbf{HGN}~\cite{kdd/MaKL19HGN}, \textbf{SASRec}~\cite{KangM18SASRec}, \textbf{BERT4Rec}~\cite{SunLWPLOJ19BERT4Rec}, \textbf{FDSA}~\cite{ZhangZLSXWLZ19FDSA}, and \textbf{S$^3$Rec}~\cite{ZhouWZZWZWW20S3Rec}. 
We adopt eight state-of-the-art generative recommenders: \textbf{P5-SID}, \textbf{P5-CID}, \textbf{P5-SemID}~\cite{HuaXGZ23P5Howtoindex}, \textbf{TIGER}~\cite{RajputMSKVHHT0S23TIGER},  \textbf{IDGenRec}~\cite{Tan24IDGenRec}, 
\textbf{LETTER}~\cite{cikm24LETTER}, 
\textbf{ELMRec}~\cite{WangCFS24ELMRec}, and \textbf{LC-Rec}~\cite{ZhengHLCZCW24LCRec}. All results are averaged over three runs with different seeds. Further details are provided in Appendix~\ref{sec:app_baseline}.

\vspace{1mm}
\noindent
\textbf{Implementation Details}. 
We implemented \ours~on OpenP5~\cite{sigir/XuHZ24OpenP5}. The model is initialized with T5-small~\cite{JMLR/Raffel2020/T5} to be consistent with existing works~\cite{HuaXGZ23P5Howtoindex, WangCFS24ELMRec, Tan24IDGenRec}. The model was trained with the Adam~\cite{KingmaB14Adam} optimizer with a learning rate of 0.001 and a linear scheduler with a warm-up ratio of 0.05. We set the maximum text length to 128 and the batch size to 128. For hierarchical IDs, $l$ and $c$ are tuned among $\{5, 7, 9\}$ and $\{32, 128, 512\}$. We tuned $k$ for collaborative semantics in $\{5, 10, 20\}$. We use a constrained beam search with a beam size of 50. The maximum number of items in sequence was set to 20 following \citet{ZhengHLCZCW24LCRec}. More details are in Appendix~\ref{sec:app_setup}.
For fair comparison, we carefully modified some experimental settings in baselines~\cite{WangCFS24ELMRec, Tan24IDGenRec}, with details provided in Appendix~\ref{sec:app_idgenrec} and~\ref{sec:app_elmrec}.

\section{Experimental Results}\label{sec:results}

\begin{table}[]\footnotesize
\vspace{-1.5mm}
\setlength{\tabcolsep}{3.82pt}
\renewcommand{\arraystretch}{0.9}
\begin{tabular}{c|cc|cc}
\toprule
\multirow{2}{*}{Model}    & \multicolumn{2}{c|}{Beauty} & \multicolumn{2}{c}{Toys} \\
                          & R@5          & N@5         & R@5         & N@5       \\ \midrule
\textbf{G{\scriptsize RAM}} & \textbf{0.0641} & \textbf{0.0451} & \textbf{0.0718} & \textbf{0.0516}  \\ \midrule
w/o hierarchy & 0.0605 & 0.0438 & 0.0630 & 0.0466 \\ 
w/o CF ($a_{CF}$) & 0.0567 & 0.0396 & 0.0589 & 0.0406 \\
\midrule
w/o user prompt ($T_u$) & 0.0634 & 0.0443 & 0.0709 & 0.0510 \\
w/o item prompt ($T_i$) & 0.0582 & 0.0404 & 0.0574 & 0.0397 \\
\midrule
w/o linking ($a_{ID}$) & 0.0628 & 0.0441 & 0.0702 & 0.0507 \\
w/o position ($\mathbf{P}$) & 0.0563 & 0.0395 & 0.0665 & 0.0465 \\
\bottomrule
\end{tabular}
\caption{Ablation study of \ours. We examined the effect of (i) semantic-to-lexical translation, (ii) the multi-granular prompts, and (iii) additional techniques. }\label{tab:exp_ablation}
\end{table}

\subsection{Overall Performance}

We thoroughly evaluate \ours's effectiveness on four real-world datasets, as presented in Table~\ref{tab:exp_overall}. Our key findings are as follows. 

(i) \ours~consistently outperforms the state-of-the-art models, achieving up to 16.0\% and 13.6\% improvement in R@5 and N@5. Compared to the best generative models (LC-Rec and IDGenRec), \ours~shows remarkable gains of up to 32.3\% and 34.1\% in R@5 and N@5. It indicates the effectiveness of capturing user preferences through item relationships and rich item information.

(ii) The generative recommendation models incorporating hierarchical item relationships into IDs (P5-SemID, LC-Rec, and \ours) demonstrate strong performance, highlighting the importance of hierarchical structures in recommendation. \ours~enhances this advantage by mapping these relationships to LLM vocabulary tokens, enabling better utilization of pre-trained language understanding capabilities. 

(iii) \ours~consistently outperforms IDGenRec using textual IDs, yielding average gains of 46.3\% in R@5 and 45.9\% in N@5. This is due to the inherent limitation of concise item IDs, which compress item information and lead to information loss. In contrast, \ours's late fusion approach preserves comprehensive information by delaying information aggregation until the decoder.

\subsection{Ablation Study}

We analyze the contributions of key components in \ours, as shown in Table~\ref{tab:exp_ablation} (See also Appendix~\ref{sec:app_synergy} and~\ref{sec:app_backbone} for additional results).

\noindent
\textbf{Semantic-to-Lexical Translation}. Both hierarchical and collaborative semantics yield 27.2\% and 10.8\% improvement in N@5, respectively. `w/o hierarchy' denotes selecting representative tokens from each item text without hierarchy. Hierarchical clustering is particularly effective, demonstrating the benefits of considering item relationships beyond simple vocabulary tokens. Our analysis also reveals that collaborative patterns are successfully integrated into language semantics. We also investigate the generalizability of semantic-to-lexical translation in Appendix~\ref{sec:app_cross}.

\noindent
\textbf{Multi-granular Prompts}. 
Both user prompts and item prompts contribute to accuracy, improving N@5 by up to 1.2\% and 30.1\%, respectively. The substantial gain from item prompts highlights the role of expressing detailed information. User prompts are particularly effective in preserving sequential information, yielding higher gains for longer sequences, as evidenced in Appendix~\ref{sec:app_userprompt}.

\noindent
\textbf{Additional Techniques}. Information linking contributes to seamless late fusion, boosting N@5 by up to 1.8\%. The multi-granular information is bridged beyond the length barrier of texts with the simple technique. The position embedding improves N@5 by up to 13.2\%, making LLMs distinguish the sequential order of items well during late fusion. It exhibits that the item orders in a sequence play a pivotal role in predicting the next items.

\begin{figure}[t]\small
\centering
\includegraphics[width=1.0\linewidth]{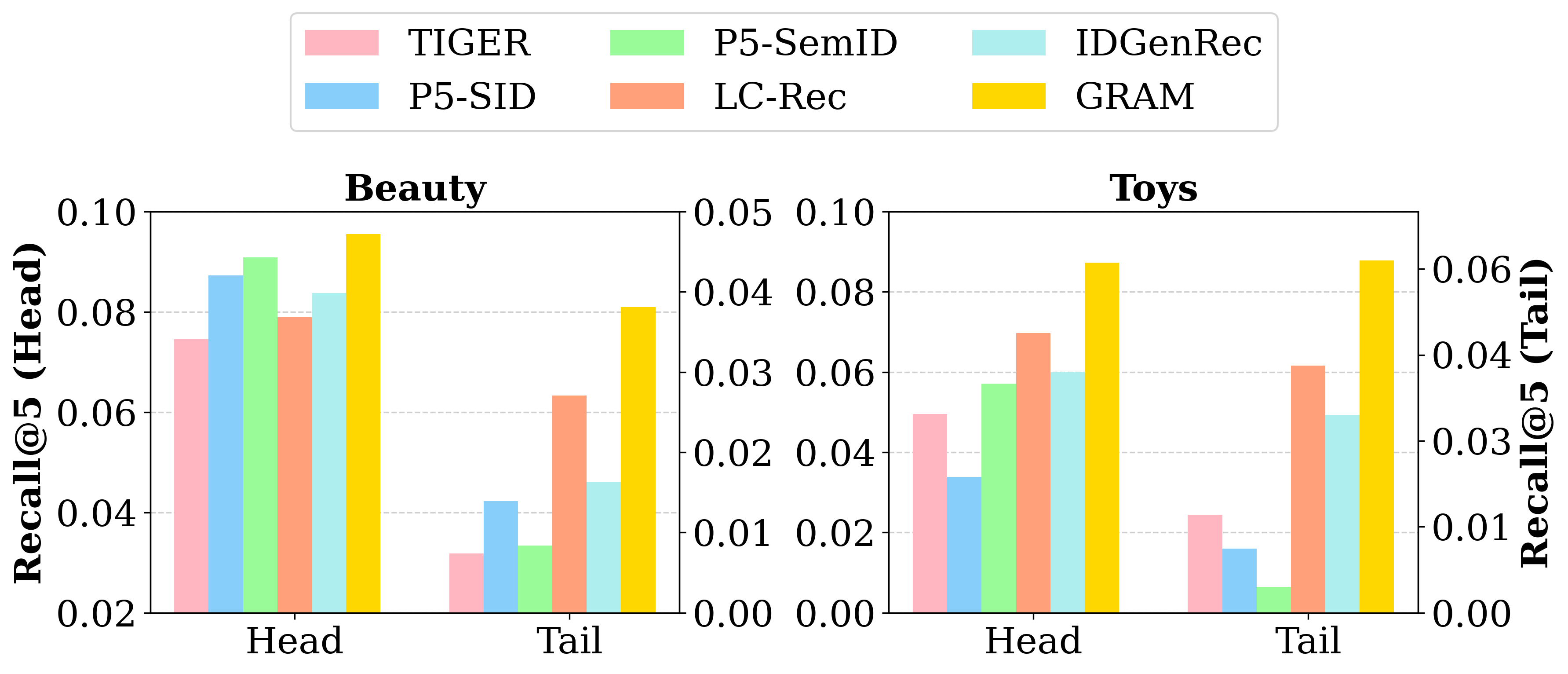}
\includegraphics[width=1.0\linewidth]{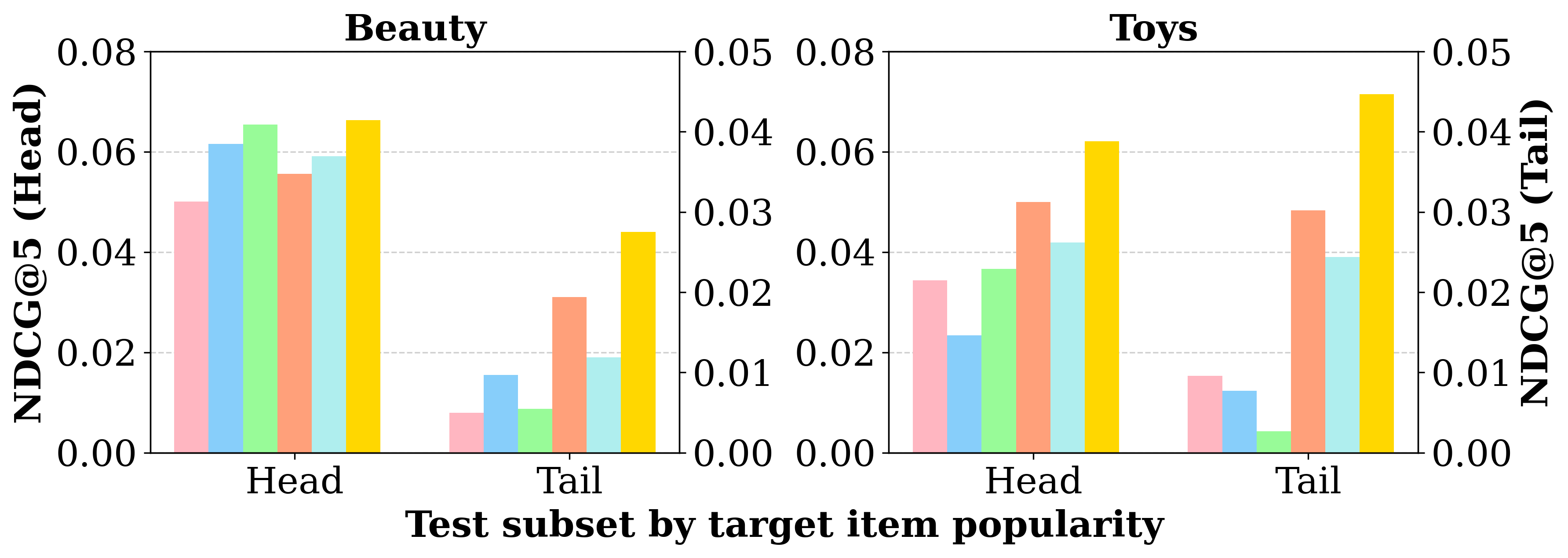}
\caption{Performance of generative recommendation models depending on the popularity of target items.}\label{fig:exp_popularity}
\end{figure}

\subsection{In-depth Analysis}
\begin{table}[]\footnotesize
\centering
\renewcommand{\arraystretch}{1.0}
\vspace{-1.5mm}
\begin{tabular}{l|cc|cc}
\toprule
\multirow{2}{*}{ID type} & \multicolumn{2}{c|}{Beauty} & \multicolumn{2}{c}{Toys} \\
 & R@5 & N@5 & R@5 & N@5 \\ \midrule
\textbf{Hierarchical ID}& \textbf{0.0641} & \textbf{0.0451} & \textbf{0.0718} & \textbf{0.0516}\\ \midrule
Title ID & 0.0478 & 0.0342 & 0.0564 & 0.0412 \\
Category ID & 0.0512 & 0.0367 & 0.0465 & 0.0350 \\
Keyword ID & 0.0605 & 0.0438 & 0.0630 & 0.0466 \\ 
RQ-VAE ID & 0.0605 & 0.0432 & 0.0662 & 0.0477 \\
\bottomrule
\end{tabular}
\caption{Performance of \ours~over various IDs.}\label{tab:exp_id}
\vspace{-2.5mm}
\end{table}

\noindent
\textbf{Effect of Hierarchical ID}.
Table~\ref{tab:exp_id} shows the effectiveness of hierarchical IDs compared to various IDs: Title ID, Category ID\footnote{For items with the same categories or titles, we append additional digits to ensure uniqueness.}, Keyword ID (extracting keywords from item metadata without clustering), and RQ-VAE ID~\cite{RajputMSKVHHT0S23TIGER, ZhengHLCZCW24LCRec}. (i) Hierarchical ID outperforms Title and Category ID by up to 31.7\% and 47.7\% in N@5, showing raw metadata lacks sufficient granularity. (ii) Compared to Keyword ID, hierarchical ID shows up to 10.8\% gains in N@5, highlighting the benefits of capturing hierarchical relationships. (iii) Hierarchical ID improves N@5 by 8.2\% over RQ-VAE ID, demonstrating the benefits of using LLM vocabulary instead of newly defined tokens that may create semantic gaps.

\noindent
\textbf{Performance on Head/Tail Items}.
As shown in Figure~\ref{fig:exp_popularity}, we analyze the performance of \ours~and generative models depending on the popularity of target items by splitting the entire user sequence into Head and Tail.\footnote{Head and Tail denote user groups where the target item is in the top 20\% and bottom 80\% of popularity, respectively. Refer to Appendix~\ref{sec:app_setup_headtail} for detailed statistics.} \ours~exhibits gains up to 42.6\% and 47.8\% in R@5 and N@5 for tail items compared to the best competitive method, \ie, LC-Rec. \ours~also improves performance in Head groups, boosting the performance by up to 25.3\% and 24.2\% in R@5 and N@5.

\begin{figure}[t]\small
\centering
\includegraphics[width=1.0\linewidth]{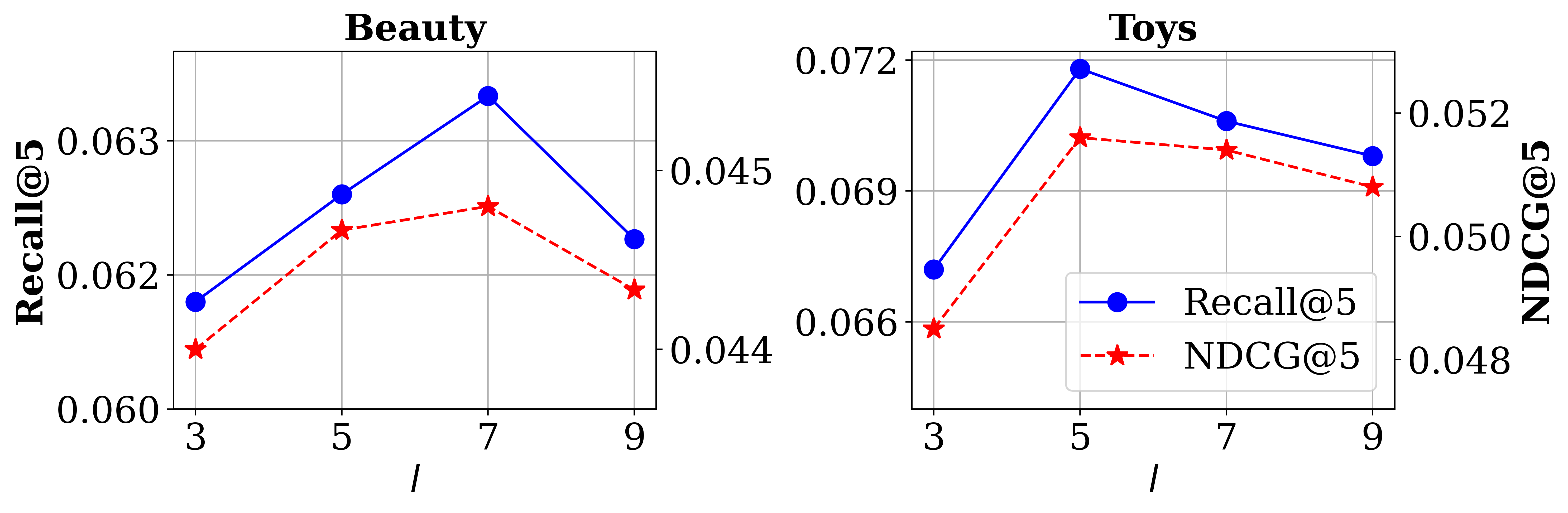}
\vspace{-3.5mm}
\caption{Performance of \ours~over varying length of identifiers $l$.}\label{fig:exp_hyper_id_length}
\end{figure}

\begin{figure}[t]\small
\centering
\includegraphics[width=1.0\linewidth]{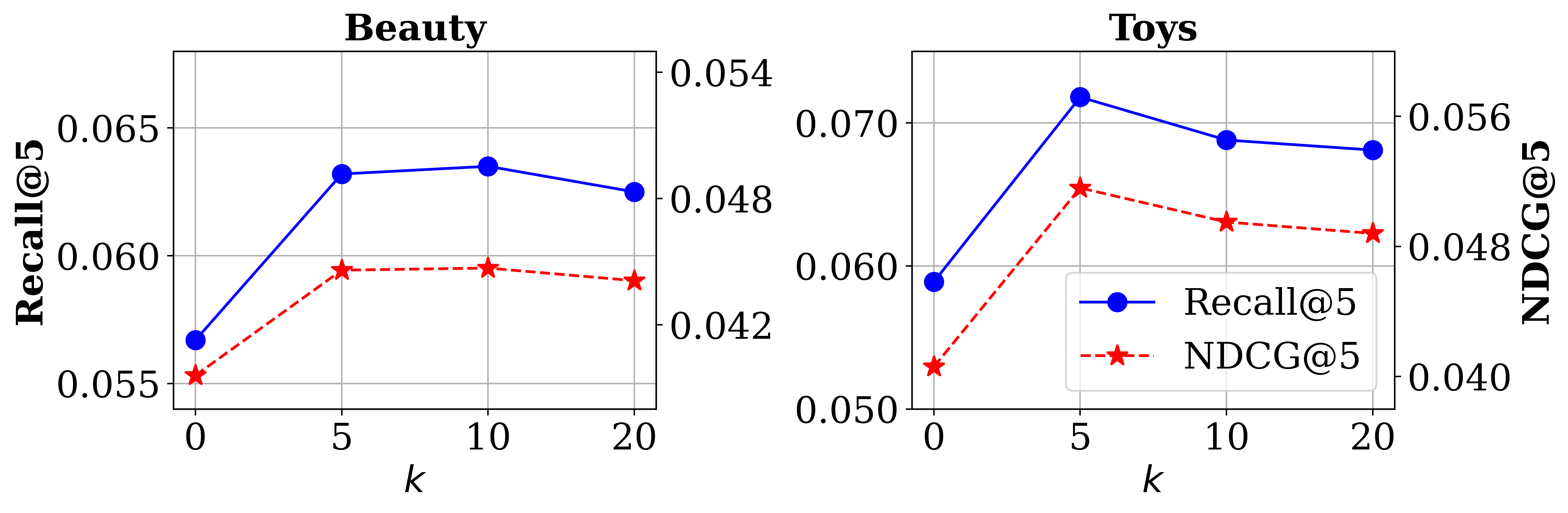}
\vspace{-3.5mm}
\caption{Performance of \ours~over varying number of the top-$k$ similar items in Eq.~\eqref{eq:top_k_fetching}.}\label{fig:exp_hyper_cf_item}
\vspace{-2.5mm}
\end{figure}

\noindent
\textbf{Hyperparameter Sensitivity}. 
Figures~\ref{fig:exp_hyper_id_length} and~\ref{fig:exp_hyper_cf_item} show the accuracy of \ours~over varying ID length $l$ and the number of the top-$k$ items in Eq.~\eqref{eq:top_k_fetching}. The optimal values for ($l$, $k$) are (5, 7) and (10, 5) for the Beauty and Toys, respectively. Collaborative semantics improves N@5 by up to 27.2\%. However, providing too many similar items introduces noise and degrades performance. An additional analysis of the cluster size $c$ is in Appendix~\ref{sec:app_results_cluster}.

\section{Conclusion}\label{sec:conclusion}
We present \ours, a novel generative recommendation model that addresses fundamental challenges in leveraging LLMs for recommendation with two key innovations: (i) semantic-to-lexical translation for bridging complex item relationships with LLMs and (ii) multi-granular late fusion for efficient and effective processing of rich item information. Our extensive experimental results across four real-world benchmark datasets validate the superiority of \ours~over existing sequential recommendation models, showing up to 16.0\% and 13.6\% gains in R@5 and N@5, respectively.

\section{Limitations}\label{sec:limitation}
While \ours~demonstrates strong performance in generative recommendation, we carefully list limitations as follows.

\noindent
\textbf{Vocabulary Selection Method}. For hierarchical semantics translation, we rely on TF-IDF scoring to select representative tokens from LLM's vocabulary space. While this provides a simple solution and serves as an efficient proof-of-concept, we conjecture that more sophisticated techniques such as well-designed neural sparse retrieval methods~\cite{sigir/FormalPC21/SPLADE, sigir/FormalLPC22/SPLADE++, cikm/ChoiLCKSL22/SpaDE} may potentially yield better representative tokens. Future work could explore integrating such approaches to improve the quality of hierarchical semantics translation while maintaining the benefits of using LLM's vocabulary.

\noindent
\textbf{Language Model Capacity}. We leverage LLMs' vocabulary and language understanding capabilities. While we demonstrate strong results using T5-small with 60M parameters and T5-base with 220M parameters as our encoder-decoder model, scaling up to a larger model (\eg, FLAN-T5-XL, T5-11B) could potentially yield even better performance, which we leave as future work.

\section*{Ethics Statement}

This work fully complies with the ACL's ethical guidelines. The scientific artifacts we have utilized are available for research under liberal licenses, and the utilization of these tools is consistent with their intended applications.

\section*{Acknowledgments} 
This work was supported by the Institute of Information \& communications Technology Planning \& Evaluation (IITP) grant and the National Research Foundation of Korea (NRF) grant funded by the Korea government (MSIT) (No. RS-2019-II190421, RS-2022-II220680, RS-2025-00564083, IITP-2025-RS-2024-00360227).

\bibliography{references}


\definecolor{userhighlight}{HTML}{FAD9D5}
\definecolor{itemhighlight}{HTML}{F2F2F2}
\definecolor{idhighlight}{HTML}{BAD8C0}
\definecolor{cfitemhighlight}{HTML}{F4D668}

\newcommand{\highlightu}[1]{\colorbox{userhighlight}{#1}}
\newcommand{\highlighti}[1]{\colorbox{itemhighlight}{#1}}
\newcommand{\highlightid}[1]{\colorbox{idhighlight}{#1}}
\newcommand{\highlightcf}[1]{\colorbox{cfitemhighlight}{#1}}
\newcommand{\highlight}[2][yellow]{\colorbox{#1}{#2}}

\begin{figure*}[htb]
\centering

\begin{tcolorbox}[
  colback=white,
  colframe=black,
  boxrule=0.5pt,
  arc=0pt,
  outer arc=0pt,
  title=Prompt example in Beauty dataset,
  fonttitle=\bfseries,
  halign=left
]\footnotesize
\label{tab:prompt}

\highlightu{\textbf{Coarse-grained user prompt $T_{u}$}}
\smallskip

What would the user purchase after ...
\highlightid{\texttt{soap-salt-sea-dead-genuine-minerals-bars}} ; \highlightid{\texttt{serum-ovi-nutrients-sea-noi-feed-ains}}?

\smallskip
\hrule
\smallskip

\highlighti{\textbf{Fine-grained item prompt $T_{i_{2}}$}}
\smallskip

\textbf{item}: \highlightid{\texttt{serum-ovi-nutrients-sea-noi-feed-ains}}; \textbf{similar items}: soap-salt-sea-dead-genuine-minerals-bar, gan-limited-33-ar-lix-treatment-eed, eye-dealing-limited-33-suitable-amp-youth, ...; 
\textbf{title}: adovia facial serum anti-aging, skin lifting, facial serum with vitamin c, dead sea salt and green tea get firmer, more radiant looking skin nonoily deep moisturizer delivers minerals amp; nutrients deep into skin satisfaction guaranteed; \textbf{brand}: na; \textbf{categories}: beauty, skincare, face, oils serums; \textbf{description}: also contains seaweed for a rich infusion of nutrients to feed and nourish your skin; \textbf{price}: 50.0; \textbf{salesrank}: beauty: 43112

\smallskip
\hrule
\smallskip
\highlighti{\textbf{Fine-grained item prompt $T_{i_{1}}$}}
\smallskip

\textbf{item}: \highlightid{\texttt{soap-salt-sea-dead-genuine-minerals-bar}}; \textbf{similar items}: serum-ovi-nutrients-sea-noi-feed-ains, eye-dealing-limited-33-suitable-amp-youth, gan-limited-33-ar-lix-treatment-eed, ...; \textbf{title}: dead sea salt deep hair conditioner for dry or damaged hair great for natural curly hair for men and women deeply hydrating and nourishing on scalp amp; hair helps to reduce dandruff and dry scalp made with pure dead sea salt and minerals, chamomile, vitamin e amp; natural aloe vera leaves your hair looking amp; feeling healthy, silky and hydrated; \textbf{brand}: na; \textbf{categories}: beauty, hair care, conditioners; \textbf{description}: made with real dead sea salt 100 pure and genuine for an infusion of 21 minerals essential to proper skin and hair function. leaves hair looking hydrated, healthy and silky. helps to reduce the appearance of dandruff and dry, flaky scalp. aloe vera deeply nourishes the scalp and hair hydrating it from within. dead sea salt contains more than 21 skin and hair rebuilding minerals such as magnesium, calcium, sulfur, bromide, iodine, sodium, zinc and potassium. we sell out fast get it soon before we run out again. we try to produce this item as fast as we can.; \textbf{price}: 19.0; \textbf{salesrank}: beauty: 7330

...
\end{tcolorbox}

\caption{Example of multi-granular prompts on the Beauty dataset for the user \texttt{A3GPKDC4PQXKFR}.}\label{fig:app_prompt}
\end{figure*}
\newpage
\newpage

\appendix

\section{Additional Related Works}\label{sec:app_relatedwork}

\noindent
\textbf{LLM-based Recommendation}.
Recent studies~\cite{recsys/DaiSZYSXS0X23, corr/abs-2311-10779DOKE, LiZC23POD, recsys24/CALRec, emnlp24d3} focused on leveraging LLMs to capture the complex semantics in detailed textual item information, providing a richer context for recommendations. Existing works are categorized into two approaches based on how they infer items~\cite{coling/24GenRecsurvey, corr/abs-2305-19860/llmrecsurvey}: \emph{discriminative} and \emph{generative} approaches. We do not cover other models that perform different recommendation tasks, \eg, yes/no (like/dislike)~\cite{acl24BinLLM, recsys23/TALLRec, abs19488CoLLM} or candidates re-ranking~\cite{corr/abs-2303-14524/ChatREC, recsys/DaiSZYSXS0X23, sigir/Liao24LLaRA}, since they are primarily beyond the scope of our work.
\begin{table*}[t]\small
\centering
\setlength{\tabcolsep}{2.4pt}
\begin{tabular}{lll}
\toprule
\textbf{Method} & \textbf{Encoding complexity} & \textbf{Decoding complexity} \\
\midrule
Early fusion & Online: $\mathcal{O}($$(|T_u|+|s|\cdot|T_{i_*}|)^2d$$+ (|T_u|+|s|\cdot|T_{i_*}|)d^2)$ & $\mathcal{O}(\tilde{i}_{|s|+1}(|T_u|+|s|\cdot|T_{i_*}|)d + \tilde{i}_{|s|+1}^2d + \tilde{i}_{|s|+1}d^2)$ \\
\midrule
\textbf{G{\scriptsize RAM}} & Online: $\mathcal{O}(|T_u|^2d + |T_u|d^2)$ & $\mathcal{O}(\tilde{i}_{|s|+1}(|T_u|+|s|\cdot|T_{i_*}|)d + \tilde{i}_{|s|+1}^2d + \tilde{i}_{|s|+1}d^2)$ \\
\textbf{(Late fusion)} & Offline: $\mathcal{O}(|s|\cdot|T_{i_*}|^2d + |s|\cdot|T_{i_*}|d^2)$ & - \\
\bottomrule
\end{tabular}
\caption{Computational complexity comparison between early and late fusion approaches. $|s|$ denotes the number of items in the user sequence. $|T_u|$ and $|T_{i_*}|$ are the number of tokens for the user prompt and the item prompt, respectively. $\tilde{i}_{|s|+1}$ the number of tokens for the target item ID.}
\label{tab:complexity}
\end{table*}

\noindent
\textbf{Discriminative Recommendation}. 
The discriminative approach employs LLMs as a sequence encoder to encode user/item text. The relevance between user and item representations is computed in a manner analogous to traditional sequential recommendation models. Several studies~\cite{HouMZLDW22UniSRec, LiWLFSSM23Recformer, LiuMXLYL0023TASTE, TangWZZZL23MIRACLE} have proposed methods to encode user and item representations using the text that comprises various item attributes, \eg, title, brand, and categories. While discriminative methods are one of the prominent pillars of text-based recommendation models, we omit further details as they fall outside the primary scope of our work.

\section{Examples of Prompts}\label{sec:app_prompt}
\noindent
Figure~\ref{fig:app_prompt} illustrates the multi-granular prompts on the Beauty dataset used as inputs for \ours. The prompts consist of a coarse-grained user prompt $T_u$ and fine-grained user prompts $T_i$ as in Eq.~\eqref{eq:fid_input}.

\section{Efficiency Analysis}\label{sec:app_efficiency}
\subsection{Theoretical Complexity Analysis}\label{sec:app_complexity}
Table~\ref{tab:complexity} shows the theoretical complexity of early and late fusion for processing user and item prompts. While early fusion concatenates all texts at the input level with quadratic complexity $\mathcal{O}((|T_u|+|s|\cdot|T_{i_*}|)^2d)$, our late fusion processes user and item prompts separately, achieving significant gains in efficiency. Specifically, \ours~reduces online computation to $\mathcal{O}(|T_u|^2d)$ by processing the item prompt $\mathcal{O}(|s|\cdot|T_{i_*}|^2d)$ in offline.

\begin{table}[]\footnotesize
\centering
\begin{tabular}{c|cc}
\toprule
Phase & \begin{tabular}[c]{@{}c@{}}Beauty\\ Time (m)\end{tabular} & \begin{tabular}[c]{@{}c@{}}Toys\\ Time (m)\end{tabular} \\ \midrule
CF model training & 2 & 2 \\
Top-k CF item retrieval & 1 & 1 \\
Text encoding & 17 & 16 \\
Hierarchical clustering & 23 & 10 \\ \midrule
Total time & 43 & 29 \\ 
\bottomrule
\end{tabular}
\caption{Preprocessing time (minutes) of \ours. Note that CF model training time can vary depending on the model, and we utilized SASRec.}\label{tab:app_preprocessing_time}
\end{table}

\begin{table}[]\footnotesize
\centering
\begin{tabular}{c|cc|cc}
\toprule
\multirow{2}{*}{Model} & \multicolumn{2}{c|}{Beauty} & \multicolumn{2}{c}{Toys} \\
 & R@5 & Time (s) & R@5 & Time (s) \\ \midrule
IDGenRec & 0.0463 & 0.1504 & 0.0462 & 0.1423 \\
LC-Rec & 0.0503 & 1.8312 & 0.0506 & 1.8125 \\ \midrule
G{\scriptsize RAM} & 0.0641 & 0.2008 & 0.0718 & 0.1605 \\
\bottomrule
\end{tabular}
\caption{Comparison of accuracy and inference time (seconds) between \ours~and key generative baselines. The inference time is measured by processing all user sequences in a single batch on a single A6000 GPU. We report the average inference time per user sequence.}\label{tab:app_inference_time}
\end{table}

For instance, with $|T_u|$=128, $|T_{i_*}|$=512, and $|s|$=20 items, early fusion requires encoding 10,368 tokens simultaneously, while \ours~only needs to encoding 128 tokens online, with the remaining encoding computations performed offline. It results in an 81$\times$ reduction (10,368 vs. 128 tokens) in online encoding complexity. Notably, since the number of tokens for target item ID $\tilde{i}_{|s|+1}$ is typically less than 10, the encoding phase dominates the computational burden~\cite{acl/23FiDO}.

\begin{table*}[]\footnotesize
\centering
\setlength{\tabcolsep}{3.4pt}
\begin{tabular}{l|cccccccccccc}
\toprule
\multicolumn{1}{c|}{\multirow{2}{*}{Model}} & \multicolumn{4}{c|}{Beauty} & \multicolumn{4}{c|}{Toys} & \multicolumn{4}{c}{Sports} \\
\multicolumn{1}{c|}{} & R@5 & N@5 & R@10 & \multicolumn{1}{c|}{N@10} & R@5 & N@5 & R@10 & \multicolumn{1}{c|}{N@10} & R@5 & N@5 & R@10 & N@10 \\ \midrule
\multicolumn{13}{l}{\textit{Reported} in~\citet{WangCFS24ELMRec}} \\ \midrule
ELMRec (original) & 0.0609 & 0.0486 & 0.0750 & \multicolumn{1}{c|}{0.0529} & 0.0713 & 0.0608 & 0.0764 & \multicolumn{1}{c|}{0.0618} & 0.0538 & 0.0453 & 0.0616 & 0.0471 \\ \midrule
\multicolumn{13}{l}{\textit{Reproduced}} \\ \midrule
ELMRec (original)  & 0.0612 & 0.0486 & 0.0759 & \multicolumn{1}{c|}{0.0533} & 0.0729 & 0.0638 & 0.0784 & \multicolumn{1}{c|}{0.0649} & 0.0503 & 0.0421 & 0.0580 & 0.0444 \\ 
\rowcolor{gray!20} 
\textbf{ELMRec (ours)}  & \textbf{0.0372} & \textbf{0.0267} & \textbf{0.0506} & \multicolumn{1}{c|}{\textbf{0.0310}} & \textbf{0.0148} & \textbf{0.0119} & \textbf{0.0193} & \multicolumn{1}{c|}{\textbf{0.0131}} & \textbf{0.0241} & \textbf{0.0181} & \textbf{0.0307} & \textbf{0.0203} \\
\bottomrule
\end{tabular}
\caption{Reproduced results of ELMRec~\cite{WangCFS24ELMRec} on the Beauty, Toys, and Sports datasets based on the indexing. We reported the result of \textbf{ELMRec (ours)} to resolve the leakage issue of sequential item IDs.}\label{tab:app_elmrec}
\end{table*}

\subsection{Empirical Efficiency Analysis}\label{sec:app_empirical_efficiency}

\noindent
\textbf{Preprocessing Phase}.
Table~\ref{tab:app_preprocessing_time} shows the time for preprocessing data before training \ours. Most importantly, hierarchical semantics indexing and collaborative semantics verbalization are \textbf{one-time} preprocessing before training. The preprocessing requires modest computational resources, taking less than an hour for the Beauty and Toys datasets. Notably, operations like clustering are highly parallelizable and can be further optimized. This preprocessing enables GRAM to achieve significant performance without affecting online inference time.

\noindent
\textbf{Inference Phase}.
Table~\ref{tab:app_inference_time} illustrates the inference time of \ours~against generative baselines. The inference time of \ours~represents worst scenarios where all computations are performed online without offline processing. \ours~achieves competitive speeds with superior performance. Even with all online processing, \ours~shows only a marginal increase in inference time ($<$50ms) compared to IDGenRec while achieving substantial performance gains. This overhead is minimal as the item prompts can be pre-computed offline. Notably, GRAM achieves 9x faster speed than LC-Rec with better performance, demonstrating its scalability for large-scale datasets.

\section{Hierarchical k-means for textual identifier}\label{sec:app_clustering}

\begin{algorithm}[H] 
\small
\caption{Hierarchical $k$-means}\label{alg:k_means}
\begin{algorithmic}
\Require Item embeddings $\mathbf{Z} = \{\mathbf{z}_1,...,\mathbf{z}_n\} \in \mathbb{R}^e$, Number of clusters $k$, Minimum cluster size $c$, Maximum depth $l$, Vocabulary $\mathcal{V}$
\Ensure Hierarchical lexical identifiers $\mathcal{H} = \{h_1,...,h_n\}$
\Function{HierarchicalClustering}{$\mathbf{Z}$, $\text{depth} = 0$}
    \State $\text{identifiers} \gets \emptyset$ 
    \If{$|\mathbf{Z}| > c$ \textbf{and} $\text{depth} < l$}
        \State $C_1,...,C_k \gets \text{$k$-means}(\mathbf{Z}, k)$ 
        \For{$i \gets 1$ \textbf{to} $k$}
            \State $t_i \gets \text{GetRepresentativeToken}(C_i, \mathcal{V})$
            \State $\mathcal{H}_i \gets \text{HierarchicalClustering}(C_i, \text{depth} + 1)$
            \State $\text{identifiers}[i] \gets \text{Concatenate}(t_i, \mathcal{H}_i)$
        \EndFor
    \EndIf
    \State \Return $\text{identifiers}$
\EndFunction
\State \
\Function{GetRepresentativeToken}{$C$, $\mathcal{V}$}
    \State $\mathbf{v} \gets \mathbf{0} \in \mathbb{R}^{|\mathcal{V}|}$
    \For{$\mathbf{z}$ \textbf{in} $C$}
        \State $\mathbf{v} \gets \mathbf{v} + \text{TfIdf}(\text{ItemText}(\mathbf{z}))$
    \EndFor
    \State \Return $\argmax_{t \in \mathcal{V}}(\mathbf{v}[t])$
\EndFunction
\end{algorithmic}
\end{algorithm}

\begin{table}[]\footnotesize
\centering
\begin{tabular}{l|cccc}
\toprule
\multicolumn{1}{c|}{Model} & R@5    & N@5    & R@10   & N@10   \\ \midrule
\multicolumn{4}{l}{\textit{Reported} in~\citet{Tan24IDGenRec}} \\ \midrule
w/ user ID    & 0.0618 & 0.0486 & 0.0814 & 0.0541 \\ \midrule
\multicolumn{4}{l}{\textit{Reproduced}} \\ \midrule
w/ user ID  & 0.0634 & 0.0487 & 0.0832 & 0.0551 \\ 
\rowcolor{gray!20} 
\textbf{w/o user ID}   & \textbf{0.0463} & \textbf{0.0328} & \textbf{0.0665} & \textbf{0.0393} \\
\bottomrule
\end{tabular}
\caption{Reproduced results of IDGenRec~\cite{Tan24IDGenRec} on the Beauty dataset based on the user IDs. We reported the result of \textbf{w/o user ID} to resolve the issue.}\label{tab:app_idgenrec}
\vspace{-3mm}
\end{table}

\section{Prompt Modification for IDGenRec}\label{sec:app_idgenrec}
\noindent
We observed a potential data leakage issue in the original implementation of IDGenRec~\cite{Tan24IDGenRec} related to the construction of user IDs. The original prompt is shown below:
\begin{tcolorbox}\small
\texttt{Considering user \{user\_id\} has interacted with items \{history\}. What is the next recommendation for the user?}
\end{tcolorbox}
\texttt{`\{user\_id\}'} takes the user ID, and \texttt{`\{history\}'} takes a sequence of concatenated item IDs generated from the item's metadata. The issue is that the user ID is constructed by concatenating all item IDs from the sequence, including validation and test items. For instance, given an item sequence $i_1 \rightarrow i_2 \rightarrow i_3 \rightarrow i_4$, according to the leave-one-out setting, $i_1 \rightarrow i_2$ is the training sequence, $i_3$ and $i_4$ are validation and test items, respectively. Here, the user ID generation process incorporates information about $i_4$, leading to data leakage during testing. To resolve this issue, we exclude user IDs in prompt sentences by following the suggestion of the authors as follows~\cite{Tan24IDGenRec}.\footnote{Please refer to the details in \url{https://github.com/agiresearch/IDGenRec/issues/1}.}

\begin{tcolorbox}\small
\texttt{Considering user has interacted with items \{history\}. What is the next recommendation for the user?}
\end{tcolorbox}
The rest of our code for IDGenRec is kept identical to the official source code provided in the paper. We also confirmed that the reported results from the original work are successfully reproduced in the original setting (`IDGenRec w/ user ID'), as shown in Table~\ref{tab:app_idgenrec}.

\section{Indexing Modification for ELMRec}\label{sec:app_elmrec}
As discussed in previous studies~\cite{RajputMSKVHHT0S23TIGER, kdd24TransRec}\footnote{Please refer to Appendix D of \citet{RajputMSKVHHT0S23TIGER} and Appendix A.6 of \citet{kdd24TransRec}.}, the original sequential indexing method in P5 has data leakage issues~\cite{Geng0FGZ22P5}. The issue arises when consecutive numeric IDs are assigned to items within user sequences, including validation and test items. For example, a sequence \texttt{[4392, 4393, ..., 4399]} where \texttt{4399} is the target item creates overlapping subword tokens (\eg, `\texttt{43}') when tokenized by SentencePiece tokenizer~\cite{Sennrich16aSentencePiece}. It creates unintended correlations between training and evaluation data.

When reproducing P5 variants, we follow the corrected setup from~\citet{HuaXGZ23P5Howtoindex, sigir/XuHZ24OpenP5} to resolve the issue, rather than using the original P5 indexing. We apply sequential indexing only to training data while excluding validation and test items.
However, ELMRec~\cite{WangCFS24ELMRec} adopts the original P5 sequential indexing in the public codebase\footnote{\url{https://github.com/WangXFng/ELMRec}}.
For fair comparison, we modified ELMRec to use the same indexing approach as P5-SID, excluding validation and test items. We also confirmed that ELMRec's performance was successfully reproduced under the provided setting as shown in Table~\ref{tab:app_elmrec}. 

\section{Additional Experimental Setup}\label{sec:app_setup}
\subsection{Datasets}\label{sec:app_dataset}
The Amazon dataset consists of user reviews and item metadata collected from 1996 to 2014. The Yelp dataset contains user reviews and business information from 2019. We create item sequences with historical user reviews from the datasets following~\citet{sigir/XuHZ24OpenP5}. The statistics of preprocessed datasets are summarized in Table~\ref{tab:statistics}.

\begin{table}[t] \small
\centering
\vspace{-3mm}
\begin{tabular}{c|cccc}
\toprule
Dataset     & \#Users   & \#Items   & \#Inters     & Density \\
\midrule
Beauty     & 22,363   & 12,101    & 198,502         & 0.0734\%  \\
Toys         & 19,412   & 11,924    & 167,597         & 0.0724\%  \\
Sports      & 35,598   & 18,357    & 296,337         & 0.0453\%  \\
Yelp     & 30,431    & 20,033    & 316,354     & 0.0519\% \\
\bottomrule
\end{tabular}
\caption{Statistics of four benchmark datasets.}
\label{tab:statistics}
\end{table}

\begin{table}[t] \small
\centering
\vspace{-2mm}
\begin{tabular}{c|cc|cc}
\toprule
\multirow{2}{*}{Subset} & \multicolumn{2}{c|}{Beauty} & \multicolumn{2}{c}{Toys} \\
 & \#Users & \#Items & \#Users & \#Items \\ \midrule
Head & 9,908 & 2,459 & 7,764 & 2,504 \\
Tail & 12,455 & 9,642 & 11,648 & 9,420 \\ 
\bottomrule
\end{tabular}
\caption{Statistics of the Beauty and Toys datasets based on test subsets by target item popularity.}
\label{tab:statistics_head_tail}
\end{table}

\subsection{Data Statistics of Head/Tail Experiments}\label{sec:app_setup_headtail}
Table~\ref{tab:statistics_head_tail} provides statistics used in Table~\ref{fig:exp_popularity}. We provide the number of items based on the groups: Head (top 20\% by popularity) and Tail (remaining 80\%). We also report the number of users based on each target item's group.

\subsection{Baselines}\label{sec:app_baseline}
We adopt six traditional sequential models and eight generative models as follows.
\begin{itemize}[leftmargin=*,topsep=0pt,itemsep=-1ex,partopsep=1ex,parsep=1ex]
\item \textbf{GRU4Rec}~\cite{HidasiKBT15GRU4Rec} is an RNN-based model that employs GRUs to encode sequences.
\item \textbf{HGN}~\cite{kdd/MaKL19HGN} employs a hierarchical gating network to capture both long-term and short-term user interests.
\item \textbf{SASRec}~\cite{KangM18SASRec} uses the last item representation as the user representation using the uni-directional Transformer encoder.
\item \textbf{BERT4Rec}~\cite{SunLWPLOJ19BERT4Rec} utilizes a bi-directional self-attention mechanism to perform the masked item prediction task.
\item \textbf{FDSA}~\cite{ZhangZLSXWLZ19FDSA} distinguishes between feature- and item-level self-attention for modeling and integration of item attributes.
\item \textbf{S$^3$Rec}~\cite{ZhouWZZWZWW20S3Rec} adopts four auxiliary self-supervised objectives to learn the correlations among sequence items and attributes.

\item \textbf{P5-SID}~\cite{HuaXGZ23P5Howtoindex} adopts sequential indexing, which assigns numeric IDs based on the order of item appearance. 
\item \textbf{P5-CID}~\cite{HuaXGZ23P5Howtoindex} employs spectral clustering to generate numeric IDs considering co-occurrence patterns of items. 
\item \textbf{P5-SemID}~\cite{HuaXGZ23P5Howtoindex} uses item metadata, \eg, categories, for assigning numeric IDs.
\item \textbf{TIGER}~\cite{RajputMSKVHHT0S23TIGER} introduces codebook IDs based on RQ-VAE. 
\item \textbf{IDGenRec}~\cite{Tan24IDGenRec} generates textual IDs with item metadata by training ID-generator. 
\item \textbf{LETTER}~\cite{cikm24LETTER} integrates hierarchical semantics, collaborative signals, and diversity when assigning RQ-VAE IDs.
\item \textbf{ELMRec}~\cite{ChenLLMX22ICLRec} incorporates high-order relationships while utilizing soft prompt and re-ranking strategy based on numeric IDs.
\item \textbf{LC-Rec}~\cite{ZhengHLCZCW24LCRec} utilizes RQ-VAE IDs and further integrates language and collaborative semantics via multi-task learning. 
\end{itemize}

\begin{table}[t]\footnotesize
\centering
\vspace{-3mm}
\setlength{\tabcolsep}{4.5pt}
\begin{tabular}{c|cccc}
\toprule
Hyperparameters & Beauty & Toys & Sports & Yelp \\
\midrule
ID length $l$ & 7 & 5 & 7 & 9 \\
\# of clusters $k$ & 128 & 32 & 32 & 32 \\
cluster size $c$ & 128 & 32 & 32 & 32 \\
\# of similar items $k$ & 10 & 5 & 10 & 5 \\
\bottomrule
\end{tabular}
\caption{Final hyperparameters for \ours.}
\label{tab:final_hyper}
\vspace{-2mm}
\end{table}

\subsection{Implementation Details}\label{sec:app_setup}

\subsubsection{Setup for Proposed Method}\label{sec:app_setup_gelato}
For hierarchical semantics extraction, we concatenated the item textual metadata as input and used NV-Embed~\cite{abs-2405-17428/NVEmbed} as a text encoder. For hierarchical semantics translation, we transformed item texts into T5 vocabulary, where $|V|=32,100$. The model with the highest N@10 on the validation set was selected for test set evaluation. We conducted all experiments on a desktop with 2 NVIDIA RTX A6000, 512 GB memory, and 2 AMD EPYC 74F3. The final hyperparameters are in Table~\ref{tab:final_hyper}.

\begin{table*}[] \small
\begin{center}
\begin{tabular}{c|c|cc|cc|cc}
\toprule
\multirow{2}{*}{Dataset} & \multirow{2}{*}{Variants} & \multicolumn{2}{c|}{All} & \multicolumn{2}{c|}{Short ($\le10)$} & \multicolumn{2}{c}{Long ($>10$)} \\
 &  & R@5 & N@5 & R@5 & N@5 & R@5 & N@5 \\ \midrule
\multirow{3}{*}{Beauty} & \textbf{G{\scriptsize RAM}} & 0.0655 & 0.0462 & 0.0609 & 0.0425 & 0.0902 & 0.0661 \\
 & w/o user prompt & 0.0634 & 0.0443 & 0.0591 & 0.0410 & 0.0860 & 0.0618 \\ \cmidrule{2-8} 
 & Gain (\%) & 3.4 & 4.3 & 3.0 & 3.5 & 4.9 & 7.0 \\ \midrule
\multirow{3}{*}{Toys} & \textbf{G{\scriptsize RAM}} & 0.0718 & 0.0516 & 0.0728 & 0.0526 & 0.0658 & 0.0457 \\
 & w/o user prompt & 0.0709 & 0.0510 & 0.0722 & 0.0522 & 0.0630 & 0.0444 \\ \cmidrule{2-8} 
 & Gain (\%) & 1.3 & 1.2 & 0.8 & 0.9 & 4.5 & 2.9 \\ 
\bottomrule
\end{tabular}
\caption{Effectiveness of user prompts based on test subsets by lengths of user sequences. Gain measures the improvement of the proposed method over `w/o user prompt.'}
\label{tab:app_user_prompt}
\end{center}
\end{table*}

\subsubsection{Setup for Baselines}\label{sec:app_setup_baselines}
\begin{itemize}[leftmargin=*,topsep=0pt,itemsep=-1ex,partopsep=1ex,parsep=1ex]

\item  \textbf{Traditional methods}: We implemented traditional recommendation models on the open-source library RecBole~\cite{sigir/23recbole1.2}. The baselines are optimized using Adam optimizer with a learning rate of 0.001, batch size of 256, and an embedding dimension of 64. We stopped the training if the validation NDCG@10 showed no improvement for 10 consecutive epochs. For evaluation, we report test set accuracy using model checkpoints that achieved the highest validation scores. We followed the original papers' configurations for other hyperparameters and carefully tuned them if not specified.

\item \textbf{P5-variants}~\cite{HuaXGZ23P5Howtoindex}: We utilized T5-small following the original implementation provided by the authors\footnote{\url{https://github.com/Wenyueh/LLM-RecSys-ID}}. For the sequential indexing approach (P5-SID), we excluded validation and test items during the assignment of numeric IDs. (For a detailed discussion, see  Appendix~\ref{sec:app_elmrec}.)

\item \textbf{TIGER}~\cite{RajputMSKVHHT0S23TIGER}: As the official code was not publicly available, we implemented the model based on the specifications detailed in the paper. We used the Sentence-T5~\cite{acl/NiACMHCY22/SentenceT5} for extracting semantic embeddings with a hidden dimension of 768. The vocabulary size was set to 1024 (256$\times$4).
 
\item \textbf{IDGenRec}~\cite{Tan24IDGenRec}: Following the official code\footnote{\url{https://github.com/agiresearch/IDGenRec}}, we used T5-small but excluded user IDs to prevent data leakage following the authors' request, incurring lower accuracy than the original paper. (Refer to the further discussion in Appendix~\ref{sec:app_idgenrec}.) We follow the `standard recommendation' setting in the paper.

\item \textbf{LETTER}~\cite{cikm24LETTER}: We initiated LETTER on TIGER using the official code\footnote{\url{https://github.com/HonghuiBao2000/LETTER}}. We followed the original setting and adopted a 4-level codebook configuration (256$\times$4), with each codebook having a dimension of 32. The original settings were maintained, and the hyperparameters $\alpha$ and $\beta$ were thoroughly tuned within the original search range.  

\item \textbf{ELMRec}~\cite{WangCFS24ELMRec}: We used T5-small following the official codebase\footnote{\url{https://github.com/WangXFng/ELMRec}}. While the original paper utilized validation loss for checkpoint selection and early stopping, we found this approach leads to convergence issues and significantly lower performance in our experiments. Thus, we adopted validation NDCG@10 for checkpoint selection and early stopping. We only performed direct and sequential recommendation tasks for the Yelp dataset, excluding the explanation generation task. Additionally, since items in user sequences can reappear as targets in the Yelp dataset, we did not employ the proposed reranking strategy. We also thoroughly fine-tuned hyperparameters of ELMRec ($\alpha, \beta, N, \sigma, L$) following the search range of the original paper.

\item \textbf{LC-Rec}~\cite{ZhengHLCZCW24LCRec}: We followed the official code\footnote{\url{https://github.com/RUCAIBox/LC-Rec}}, including the data pre-processing described in the paper. We fully fine-tuned LLaMA-7B~\cite{TouvronLIMLLRGHARJGL23LLaMA} using the authors' hyperparameters. The only exception was that the number of samples for the `fusion sequence recommendation task' was adjusted according to each dataset's interaction numbers. Specifically, we use 15k samples for the Toys, 20k for the Beauty, and 25k for the Yelp and Sports datasets. For generating user preferences, we employed \texttt{gpt-4o-mini-2024-07-18}\footnote{\url{https://chatgpt.com}} on the Amazon datasets. The Yelp dataset was excluded due to its lack of metadata (\ie, reviews) for preference generation.

\end{itemize}

\begin{table}[]\footnotesize
\centering
\setlength{\tabcolsep}{3.0pt}
\renewcommand{\arraystretch}{1.1}
\begin{tabular}{c|cc|cc}
\toprule
\multicolumn{1}{l|}{} & \multicolumn{2}{c|}{Beauty} & \multicolumn{2}{c}{Toys} \\
\multicolumn{1}{l|}{} & R@5 & N@5 & R@5 & N@5 \\ \midrule
P5-SID & 0.0465 & 0.0329 & 0.0216 & 0.0151 \\
P5-CID & 0.0465 & 0.0325 & 0.0223 & 0.0143 \\
P5-SemID & 0.0459 & 0.0327 & 0.0264 & 0.0178 \\
\textbf{P5 + HID} & \textbf{0.0582} & \textbf{0.0404} & \textbf{0.0574} & \textbf{0.0397} \\ \midrule
IDGenRec & 0.0463 & 0.0328 & 0.0462 & 0.0323 \\
\textbf{IDGenRec + HID} & \textbf{0.0499} & \textbf{0.0352} & \textbf{0.0656} & \textbf{0.0475} \\ \bottomrule
\end{tabular}
\caption{Effectiveness of hierarchical IDs (HID) when applying to existing generative recommenders.}\label{tab:app_crossmodel_hid}
\end{table}

\begin{table}[]\footnotesize
\centering
\setlength{\tabcolsep}{3.0pt}
\renewcommand{\arraystretch}{1.1}
\begin{tabular}{c|cc|cc}
\toprule
\multicolumn{1}{l|}{} & \multicolumn{2}{c|}{Beauty} & \multicolumn{2}{c}{Toys} \\
\multicolumn{1}{l|}{} & R@5 & N@5 & R@5 & N@5 \\ \midrule
P5-SID & \textbf{0.0465} & \textbf{0.0329} & 0.0216 & 0.0151 \\
\textbf{P5-SID + CF} & 0.0454 & 0.0295 & \textbf{0.0419} & \textbf{0.0263} \\ \midrule
IDGenRec & 0.0463 & 0.0328 & 0.0462 & 0.0323 \\
\textbf{IDGenRec + CF} & \textbf{0.0623} & \textbf{0.0420} & \textbf{0.0513} & \textbf{0.0337} \\ \bottomrule
\end{tabular}
\caption{Effectiveness of collaborative semantics verbalization (CF) when applying to existing generative recommenders.}\label{tab:app_crossmodel_cf}
\vspace{-2.5mm}
\end{table}

\section{Additional Experimental Results}\label{sec:app_results}

\subsection{In-depth Study of User Prompts}\label{sec:app_userprompt}
Table~\ref{tab:app_user_prompt} shows the impact of user prompts across different sequence lengths to understand their role in the recommendation. While \ours~consistently outperforms the variant without user prompts across all groups, the improvements are particularly significant for long sequences (length > 10). For the Beauty dataset, incorporating user prompts achieves gains of up to 7.0\% in N@5 for long sequences, compared to 3.5\% for short sequences. Similar trends are observed in the Toys dataset, with gains of 2.9\% and 0.9\% in N@5 for long and short sequences, respectively.
It demonstrates that user prompts effectively preserve sequential dependencies, especially for longer user histories, without sacrificing efficiency through late fusion.

\subsection{Generalizability Analysis of Components}\label{sec:app_cross}
As shown in Table~\ref{tab:app_crossmodel_hid}, hierarchical IDs demonstrate consistent improvements when applied to existing generative recommendation models. When applied to P5, it yields substantial gains of up to 23.0\% in N@5. Similarly, IDGenRec achieves improvements of up to 47.1\% in and N@5. It implies that semantic-to-lexical translation effectively bridges the gap between item relationships and language understanding capabilities of LLMs, even with other generative recommendation models. 

\begin{table}[] \footnotesize
\begin{center}
\centering
\setlength{\tabcolsep}{2.9pt}
\begin{tabular}{c|c|cc|cc}

\toprule
\multirow{2}{*}{Semantic} & \multirow{2}{*}{Fusion} & \multicolumn{2}{c|}{Beauty} & \multicolumn{2}{c}{Toys} \\
 &  & R@5 & N@5 & R@5 & N@5 \\ \midrule
\cmark & \cmark & \textbf{0.0641} &\textbf{0.0451} & \textbf{0.0718} & \textbf{0.0516} \\
\cmark & \xmark & 0.0582 & 0.0404 & 0.0574 & 0.0397 \\ 
\xmark & \cmark & 0.0534 & 0.0382 & 0.0538 & 0.0384 \\
\xmark & \xmark & 0.0516 & 0.0366 & 0.0459 & 0.0336 \\ 
\bottomrule
\end{tabular}
\caption{Performance of \ours~based on two main components: semantic-to-lexical translation (`Semantic') and multi-granular late fusion (`Fusion').}
\label{tab:app_synergy}
\end{center}
\end{table}

\begin{table}[]\footnotesize
\centering
\setlength{\tabcolsep}{3.0pt}
\renewcommand{\arraystretch}{1.1}
\begin{tabular}{c|cc|cc}
\toprule
\multicolumn{1}{l|}{} & \multicolumn{2}{c|}{Beauty} & \multicolumn{2}{c}{Toys} \\
\multicolumn{1}{l|}{} & R@5 & N@5 & R@5 & N@5 \\ \midrule
T5-small & 0.0641 & 0.0451 & 0.0718 & 0.0516 \\
T5-base & \textbf{0.0646} & \textbf{0.0457} & \textbf{0.0719} & \textbf{0.0520} \\
\bottomrule
\end{tabular}
\caption{Performance of \ours~over various model sizes. We adopt `T5-small' for other experiments.}\label{tab:app_model_size}
\end{table}

Table~\ref{tab:app_crossmodel_cf} illustrates the impact of collaborative semantics verbalization. For compatibility, collaborative signals are appended after the original prompt sentence of each model. Notably, IDGenRec shows gains of up to 34.6\% in R@5 with collaborative signals. However, the improvements vary by model, indicating that separating item prompts is more beneficial than direct concatenation with existing prompts. This aligns with our multi-granular design principle of maintaining distinct granularities of information.

\begin{table*}[]\scriptsize
\centering
\setlength{\tabcolsep}{3.32pt}
\renewcommand{\arraystretch}{1.2}

\begin{tabular}{ccc|cc|cc|cc}
\toprule
\multicolumn{1}{c|}{\multirow{2}{*}{Model}} & \multicolumn{2}{c|}{Beauty} & \multicolumn{2}{c|}{Toys} & \multicolumn{2}{c|}{Sports} & \multicolumn{2}{c}{Yelp} \\
\multicolumn{1}{c|}{} & R@20 & N@20 & R@20 & N@20 & R@20 & N@20 & R@20 & N@20 \\ \midrule
\multicolumn{9}{c}{\textit{Traditional recommendation models}} \\ \midrule
\multicolumn{1}{c|}{GRU4Rec} & 0.0934 & 0.0430 & 0.0794 & 0.0372 & 0.0573 & 0.0247 & 0.0659 & 0.0273 \\
\multicolumn{1}{c|}{HGN}  & 0.0903 & 0.0372 & 0.0840 & 0.0352 & 0.0539 & 0.0221 & 0.0777 & \ul{0.0366} \\
\multicolumn{1}{c|}{SASRec} & 0.0676 & 0.0300 & 0.0562 & 0.0271 & 0.0315 & 0.0137 & 0.0457 & 0.0275 \\
\multicolumn{1}{c|}{BERT4Rec} & 0.0729 & 0.0294 & 0.0567 & 0.0231 & 0.0384 & 0.0154 & 0.0655 & 0.0274 \\
\multicolumn{1}{c|}{FDSA} & \ul{0.1065} & \ul{0.0551} & \ul{0.1057} & \ul{0.0578} & 0.0568 & 0.0281 & \ul{0.0859} & 0.0365 \\
\multicolumn{1}{c|}{S$^3$Rec}  & 0.0984 & 0.0405 & 0.0927 & 0.0388 & 0.0577 & 0.0243 & 0.0533 & 0.0212 \\ \midrule
\multicolumn{9}{c}{\textit{Generative recommendation models}} \\ \midrule
\multicolumn{1}{c|}{P5-SID} & 0.0823 & 0.0431 & 0.0407 & 0.0208 & 0.0518 & 0.0276 & 0.0625 & 0.0302 \\
\multicolumn{1}{c|}{P5-CID} & 0.0928 & 0.0456 & 0.0555 & 0.0236 & 0.0597 & 0.0299 & 0.0567 & 0.0250 \\
\multicolumn{1}{c|}{P5-SemID} & 0.0958 & 0.0468 & 0.0619 & 0.0278 & \ul{0.0675} & \ul{0.0342} & 0.0504 & 0.0225 \\
\multicolumn{1}{c|}{TIGER} & 0.0775 & 0.0355 & 0.0657 & 0.0282 & 0.0418 & 0.0179 & 0.0406 & 0.0171 \\
\multicolumn{1}{c|}{IDGenRec$^\dagger$} & 0.0930 & 0.0460 & 0.0899 & 0.0446 & 0.0579 & 0.0273 & 0.0636 & 0.0311 \\
\multicolumn{1}{c|}{ELMRec$^\dagger$} & 0.0444 & 0.0231 & 0.0114 & 0.0067 & 0.0263 & 0.0157 & 0.0367 & 0.0225 \\
\multicolumn{1}{c|}{LC-Rec} & 0.0973 & 0.0485 & 0.0964 & 0.0485 & 0.0550 & 0.0257 & 0.0720 & 0.0341 \\
\midrule
\rowcolor{gray!20} 
\multicolumn{1}{c|}{\textbf{G{\scriptsize RAM}}}  & \textbf{0.1216} & \textbf{0.0613} & \textbf{0.1303} & \textbf{0.0682} & \textbf{0.0796} & \textbf{0.0375} & \textbf{0.1012} & \textbf{0.0476} \\ \midrule
\multicolumn{1}{c|}{Gain (\%)}  & 14.2$^*$ & 11.3$^*$ & 23.3$^*$ & 18.0$^*$ & 18.0$^*$ & 9.8$^*$ & 17.8$^*$ & 30.1$^*$ \\
\bottomrule
\end{tabular}
\caption{Performance comparison with cutoff 20. The best model is marked in \textbf{bold}, and the second-best model is \ul{underlined}. Gain measures the improvement of the proposed method over the best competitive baseline. `$*$' indicates statistical significance $(p < 0.05)$  by a paired $t$-test.}\label{tab:exp_overall_cutoff20}
\end{table*}

\begin{figure}[t]\small
\centering
\includegraphics[width=1.0\linewidth]{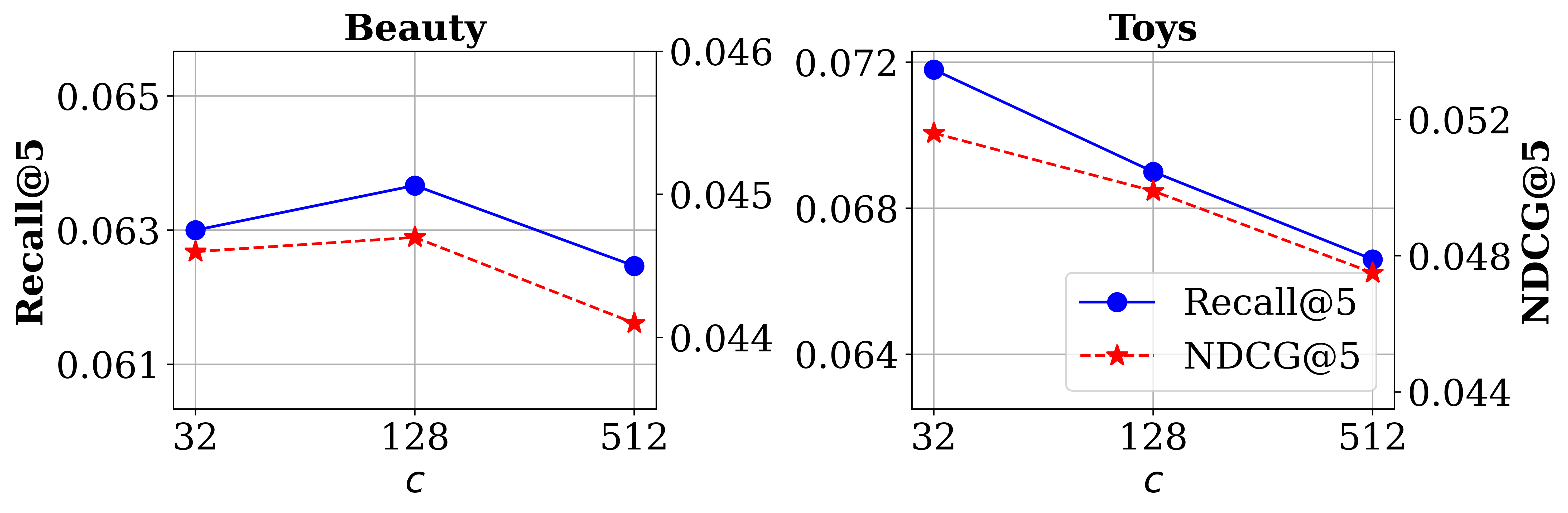}
\vspace{-6.5mm}
\caption{Performance of \ours~over varying numbers of cluster size $c$ for hierarchical clustering.}\label{fig:exp_hyper_cluster_num}
\vspace{-3mm}
\end{figure}

\begin{figure*}
\centering
\includegraphics[width=1.0\linewidth]{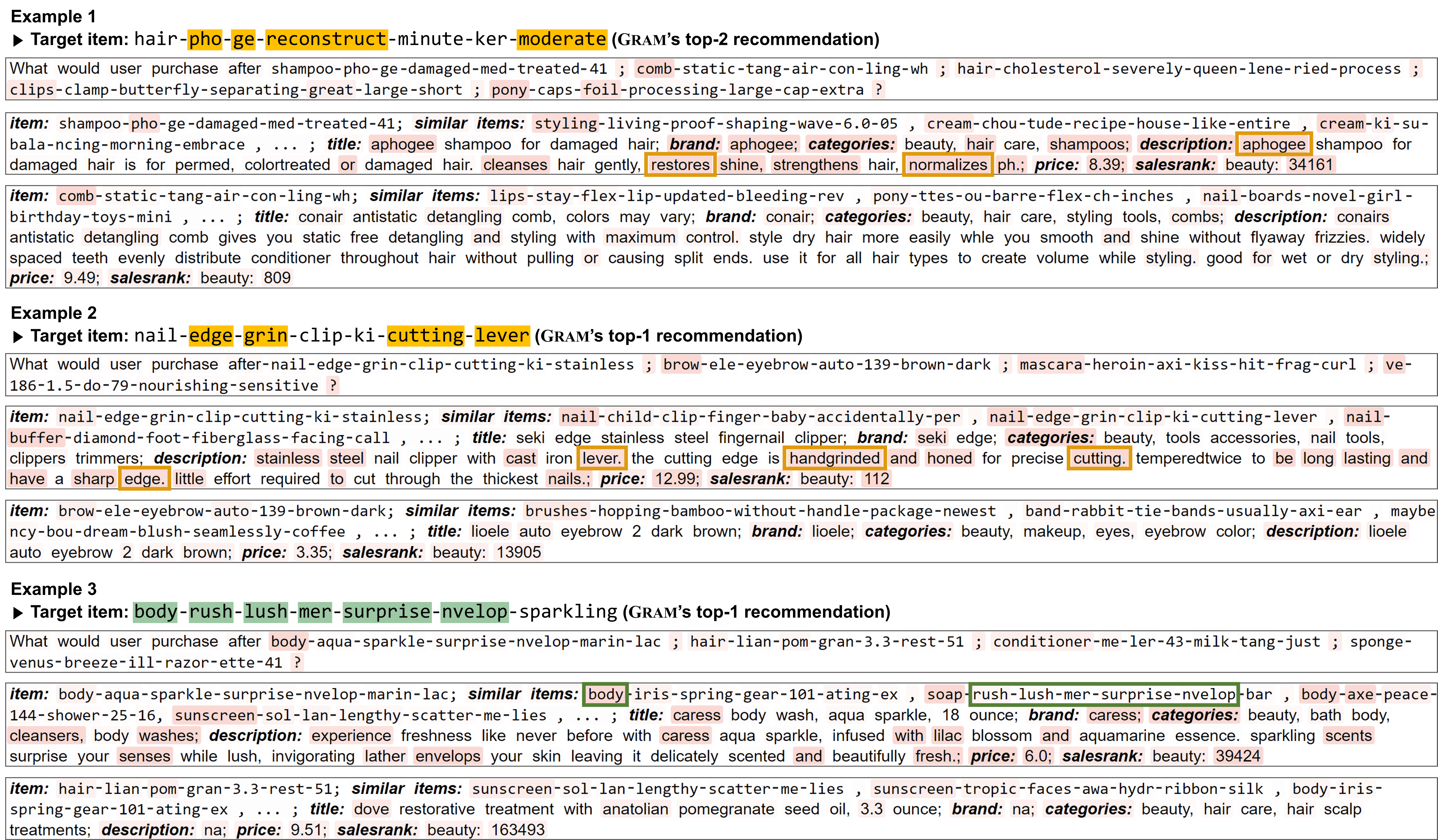}
\vspace{-4.5mm}
\caption{Visualization of cross-attention scores in \ours. Our model utilizes not only item IDs but also similar items and detailed textual information for recommendations. Darker shades of red indicate higher attention scores. The orange box highlights textual descriptions, while the green box highlights the collaborative signals involved in the recommendation.}\label{fig:qualitative}
\end{figure*}

\subsection{Synergistic Effects between Components}\label{sec:app_synergy}
To analyze the effectiveness of key components of \ours, we extend our investigation beyond the ablation study in Table~\ref{tab:exp_ablation}. Table~\ref{tab:app_synergy} demonstrates both the individual and combined effects of the components. Each component independently improves performance over \ours~without any components (fourth row), with semantic-to-lexical translation showing a more substantial impact compared to multi-granular late fusion.
Most importantly, the results reveal significant synergistic effects between the two key components. While semantic-to-lexical translation alone yields up to 25.1\% improvement in R@5 (third and fourth row), its combination with multi-granular late fusion leads to a 33.4\% improvement (first and second row). This enhanced performance demonstrates the potential synergy of the components within our framework.

\subsection{Effect of Model Size}\label{sec:app_backbone}
To evaluate the generalizability of \ours, we examine the accuracy across different model sizes, as presented in Table~\ref{tab:app_model_size}. \ours~achieves superior performance compared to competitive baselines with both T5-small and T5-base architectures. Although increasing the model size yields modest performance improvements, we hypothesize that more extensive hyperparameter tuning could potentially achieve better performance with larger models.

\subsection{Additional Results with Cutoff 20}\label{sec:app_cutoff20}
Table~\ref{tab:exp_overall_cutoff20} shows the effectiveness of \ours~with a larger recommendation list (cutoff=20). \ours~consistently outperforms all baselines across datasets, achieving gains of up to 23.3\% and 30.1\% in R@20 and N@20, respectively. The performance gains are more pronounced than the results with the cutoff of 10, particularly for Yelp and Toys datasets. It suggests that \ours~effectively maintains accuracy in longer recommendation lists, likely due to its capacity to leverage both item relationships and comprehensive item information.

\subsection{Effect of Cluster Size}\label{sec:app_results_cluster}
Figure~\ref{fig:exp_hyper_cluster_num} demonstrates the impact of the cluster size $c$. The optimal cluster size varies across datasets. It underscores the importance of careful hyperparameter tuning when applying the model to datasets with distinct characteristics.

\subsection{Case Study}\label{sec:app_additional_case}
Figure~\ref{fig:qualitative} shows the cross-attention scores for prompts, consisting of coarse-grained user prompts and fine-grained item prompts, as defined in Eq.~\eqref{eq:fid_encoder_output}. These scores directly affect recommendation results by identifying salient tokens during the decoding process. As highlighted in red, \ours~leverages descriptions from fine-grained item prompts for recommendations. While previous methods struggle with input length limits and efficiency constraints, \ours~effectively detours the challenges through separate prompt encoding and late fusion. 

In the first example, \ours~assigns higher scores to terms such as ``\textit{restores}'' and ``\textit{normalizes}'' in the description, which captures fine-grained information of the first item. These terms semantically align with the hierarchical identifier of the target item, ``\textit{hair-pho-ge-reconstruct-minute-ker-moderate}.'' The absence of this information in other attributes underscores the importance of maintaining diverse and detailed attributes.

The third example illustrates how \ours~assigns higher scores to collaborative knowledge, specifically for similar items. Interestingly, when the CF model predicts the target item as a similar item, and \ours~effectively leverages this information to rank the target item as the top-1 recommendation. It demonstrates the critical role of collaborative semantics during prediction, effectively enhancing the ability to utilize collaborative knowledge.

\end{document}